\documentclass[preprint,noshowkeys,amsmath,amssymb,superscriptaddress,nofootinbib,longbibliography]{revtex4-1}

\usepackage{epstopdf}
\usepackage[bookmarks=true,bookmarksnumbered=true,bookmarkstype=toc]{hyperref} 
\usepackage{units}

\allowdisplaybreaks[3]
\usepackage{mathtools}
\usepackage{etoolbox}

\makeatletter
\pretocmd{\frontmatter@abstractheading}
  {\par\vspace*{5mm}}
  {}{}
\makeatother

\newcommand{\nn}{\nonumber}
\renewcommand{\Re}{\mathrm{Re}}
\renewcommand{\Im}{\mathrm{Im}}

\begin{document}
\preprint{KANAZAWA-26-09}
\preprint{UME-PP-031}

\title{Probing CP violation in dark scalar decays\\ at the SHiP experiment}

\author{Mayumi Aoki}
\email{mayumi.aoki@staff.kanazawa-u.ac.jp}
\affiliation{Institute  for  Theoretical  Physics,  Kanazawa  University,  Kanazawa  920-1192,  Japan}

\author{Takashi Shimomura}
\email{shimomura@cc.miyazaki-u.ac.jp}
\affiliation{Faculty of Education, Miyazaki University, Miyazaki, 889-2192, Japan} 

\begin{abstract}
We investigate CP violation in the decays of a GeV-scale dark scalar boson 
into mesons, muons and photons. Both scalar and pseudoscalar couplings to 
the SM fermions are considered to study how the CP structure of the dark scalar 
boson can be probed through its decays. The scalar and pseudoscalar interactions 
predominantly induce decays into two-meson and three-meson final states, respectively. 
We propose the simultaneous observation of these two classes of decay modes 
as a probe of CP violation in the dark scalar sector. The expected signal 
rates at the SHiP and FASER2 experiments are evaluated based on the dark 
scalar production from heavy-meson decays. We find that the ratio of three-meson 
to two-meson decay events at SHiP has the potential to probe a wide range of 
the CP-violating parameter space for dark scalar masses around the GeV scale. 
We also show that the two-muon and di-photon channels can provide complementary 
information for the CP violation search. 
\end{abstract}

\date{\today}
\maketitle

\section{Introduction}
The discovery of CP violation has provided an important window into the flavor 
structure of the Standard Model (SM). Following the observation of CP violation 
in neutral kaon decays, flavor changing CP violation was successfully described 
by the Cabibbo--Kobayashi--Maskawa (CKM) mechanism~\cite{Cabibbo:1963yz,Kobayashi:1973fv}. 
Nevertheless, many open questions in particle physics suggest the existence 
of additional sources of CP violation beyond the SM. One of the best-known 
examples is the origin of the baryon asymmetry of the Universe, whose explanation 
generally requires CP-violating interactions beyond those present in the SM.

One possible origin of such additional CP-violating interactions is a hidden 
or dark sector containing new scalar degrees of freedom. Gauge-singlet scalar 
bosons are among the simplest extensions of the SM. If their couplings to 
the SM particles are sufficiently small, GeV-scale dark scalar bosons naturally 
become long-lived particles (LLPs), making dedicated LLP experiments particularly 
suitable for their discovery.

An especially interesting question is whether a new dark scalar boson has 
a definite CP quantum number. A neutral spin-zero particle may be CP-even, 
CP-odd, or, if CP is violated, a CP-mixed state. Identifying such a CP-mixed 
state would provide direct evidence that the dark scalar sector contains 
a nontrivial CP structure. 

There are several ways to search for CP violation in the dark sector. 
One is to compare the decay rate of a dark scalar boson into a given final 
state with that into the corresponding CP-conjugate state. Another possibility 
is to measure angular distributions of CP-odd observables, such as 
triple products constructed from final-state particle momenta, in decays 
or productions of the dark scalars. Apart from such direct searches, one can 
also measure CP-odd observables involving SM particles, such as electric 
dipole moments (EDMs). However, all of these approaches can become challenging 
when the couplings of the dark scalar boson to SM particles are sufficiently small. 
Furthermore, measurements of angular correlations in the dark scalar decays 
will require sufficiently large event samples and detectors with good angular 
resolution. An alternative possibility is to infer the CP properties of 
the dark scalar boson from the CP quantum numbers of its decay final states, 
following the basic strategy that played a crucial role in establishing 
CP violation in neutral kaon decays. We pursue this approach in this paper.

We describe the interaction of the dark scalar $\phi$ with SM fermions $f$ 
by the effective Lagrangian 
${\cal L}_{\rm int} = (m_f/v) \bar f (\xi_f^s+i\xi_f^p\gamma_5) f \phi$,
where $\xi_f^s$ and $\xi_f^p$ denote the scalar and pseudoscalar couplings, 
respectively. Interactions of this form have also been studied in the context 
of electroweak baryogenesis~\cite{Espinosa:2011eu,Cline:2021iff}. 
In the GeV mass range, however, the same interaction leads to distinctive 
hadronic decay patterns: a CP-even scalar predominantly decays into two-meson 
final states, whereas a CP-odd scalar mainly decays into three-meson final 
states. 
Therefore, the observation of only one class of decay modes would be 
consistent with $\phi$ being a CP eigenstate, while the simultaneous observation 
of both two-meson and three-meson channels would imply both $\xi_f^s$ and 
$\xi_f^p$ to be nonzero. Such an observation would establish that the dark 
scalar is a CP-mixed state and would provide a direct experimental probe of 
CP violation in the dark sector. 
In the GeV mass range, sufficiently small couplings naturally lead to macroscopic decay lengths,
making the dark scalar an LLP candidate. 
Several experiments have been conducted to search for the LLP candidates such as MoEDAL \cite{MoEDAL:2009jwa}, FASER \cite{FASER:2022hcn}, SND \cite{SNDLHC:2022ihg},
milliQan \cite{Haas:2014dda}, and proposed or under construction such as CODEX-b \cite{CODEX-b:2019jve}, MATHUSLA \cite{Curtin:2018mvb}, ANUBIS \cite{ANUBIS:2025sgg}, FACET \cite{Cerci:2021nlb}, FASER2 (FPF) \cite{FPF:2025bor}, SHIFT \cite{Niedziela:2024khw,GarciaSoto:2025ikg}, or SHiP \cite{SHiP:2015vad,SHiP:2020vbd,ShipECN3,SHiP:2025ows}.
Among these experiments, SHiP and FASER2 are expected to provide unprecedented sensitivity to weakly coupled particles with masses below a few GeV.

Many studies of searches for CP-even and CP-odd dark scalars at these experiments have been performed so far (see reviews \cite{Feng:2022inv,Kling:2022uzy,Alimena:2025kjv,SHiP:2025ows}, and references therein). 
In this paper, we investigate the prospects for probing the CP structure of a GeV-scale dark scalar
at these experiments through its hadronic decay modes. In particular, we study whether the simultaneous observation of two- and three-meson final states can probe the
coexistence of scalar and pseudoscalar interactions and thereby test whether the dark scalar is a CP-mixed state.

This paper is organized as follows. In Section \ref{sec:models}, we introduce the effective interaction Lagrangian of the dark scalar with SM particles. In Section~\ref{sec:decay-dark-scalar}, we show the decay branching ratios and decay length of the dark scalar. We also explain the production mechanisms of the dark scalar at SHiP and FASER2. In Section~\ref{sec:results}, we present our analysis of the prospects for probing CP violation in the dark scalar sector at these experiments. Finally, we summarize our results in Section~\ref{sec:summary}.

\section{Models} \label{sec:models}
We consider a dark scalar model with CP-violating interactions. We adopt an effective Lagrangian approach so that our results can be applied to a broad class of specific models. The effective interaction Lagrangian of the dark scalar boson $\phi$ is parameterized as \cite{Espinosa:2011eu,Cline:2021iff,Kling:2022uzy}\footnote{To demonstrate explicitly that our Lagrangian can be realized in a UV model, we present one concrete example in Appendix \ref{apdx:UV-complete-model}.},
\begin{align}
\mathcal{L}_{\mathrm{int}} &= 
    -\frac{m_f}{v} \overline{f}\big( \xi^s_{f} + i \xi^p_{f} \gamma_5 \big) f \phi 
    + \xi_{\gamma} \frac{\alpha_{\rm em}}{4 \pi v} \phi F_{\mu\nu} F^{\mu\nu} + \tilde{\xi}_{\gamma} \frac{\alpha_{\rm em}}{4 \pi v} \phi F_{\mu\nu} \tilde{F}^{\mu\nu} \nn \\
    &\qquad + \xi_{g} \frac{\alpha_{s}}{12 \pi v} \phi G_{\mu\nu} G^{\mu\nu} + \tilde{\xi}_{g} \frac{\alpha_{s}}{4 \pi v} \phi G_{\mu\nu} \tilde{G}^{\mu\nu}, \label{eq:lag-int}
\end{align}
where $v=246$ GeV represents the vacuum expectation value of the SM Higgs field. The SM fermions and their masses are denoted as $f$  and $m_f$, respectively. For the fermion interactions, the minimal flavor violation hypothesis \cite{DAmbrosio:2002vsn} is assumed to avoid the stringent constraints from flavor changing processes. The photon and gluon field strengths are denoted as $F_{\mu\nu}$ and $G_{\mu\nu}$, and those with the tilde are their dual fields, respectively. The electromagnetic fine-structure constant and that for the strong interaction are written as $\alpha_{\rm em}$ and $\alpha_s$. The parameters $\xi$ are modifiers relative to the corresponding SM couplings. 
The interaction with photons and gluons can be induced via loop corrections of heavy fermions. We include the charm, bottom and top quark contributions to $\xi_g~(\tilde{\xi}_g)$ and also the charged lepton ones to $\xi_\gamma~(\tilde{\xi}_\gamma)$. The one-loop-induced couplings for these interactions are given in Appendix \ref{apdx:loop-function}. 
Note that the interactions with the electroweak gauge bosons can also be induced via loop corrections. However, these contributions are irrelevant to our study due to their heavy masses. Therefore we omit these interactions in the following discussion.

For a dark scalar boson with a mass below the GeV scale, the relevant degrees of freedom are hadrons rather than quarks and gluons. The heavy quarks are integrated out, while the light quarks and gluons are confined into hadrons, predominantly light mesons in the mass range of interest. Consequently, the dark scalar decays into hadrons rather than partonic final states. At sufficiently low energies, these hadronic decays can be described within Chiral Perturbation Theory ($\chi$PT)  \cite{Rosenzweig:1979ay,DiVecchia:1980yfw,Kawarabayashi:1980dp,Kawarabayashi:1980uh,Gasser:1984gg}. As we will see below, the dark scalar mixes with pseudoscalar mesons. Hence, we denote the dark scalar as $\hat{\phi}$ to distinguish it from its mass eigenstate below. The interaction Lagrangian of the dark scalar and mesons is given by 
\begin{align}
\mathcal{L}_{\mathrm{meson}} &= \frac{B}{2} \mathrm{Tr}
        (M_q[\hat{\phi}] \Sigma + \Sigma^\dagger M_q^\dagger[\hat{\phi}]) + \mathcal{L}_{\mathrm{axial~anomaly}}, \label{eq:lagrangian-chipt}
\end{align}
where $B$ is a constant parameter, $M_q[\hat{\phi}]$ is the field-dependent quark mass and $\Sigma$ is the light meson field. The last term is the chiral anomaly term given in Appendix \ref{apdx:chiral-anomaly}.
The light meson field $\Sigma$ is expressed in a non-linear representation as 
\begin{align}
\Sigma = \frac{f_\pi}{2} \exp \left( i\frac{\sqrt{2}}{f_\pi} \pi_a \lambda^a \right), \label{eq:sigma}
\end{align}
where $f_\pi$ is the pion decay constant and $\pi_a~(a=1,\cdots,9)$ are the pseudo-Nambu-Goldstone bosons. The corresponding $3\times 3$ matrices $\lambda^a$ are normalized as $\mathrm{Tr}(\lambda^a \lambda^b) = \delta^{ab}$ and given in Appendix \ref{apdx:meson-int-masseigenstates}. Following the procedure in \cite{Gasser:1984gg}, the CP-violating interactions are incorporated into the field dependent mass $M_q[\hat{\phi}]$ as
\begin{align}
    M_q[\hat{\phi}] &=m_q + \tilde{M}_{q} \frac{\hat{\phi}}{v},\label{eq:field-dep-mass}
\end{align}
with
\begin{subequations}
\begin{align}
    m_q &= \mathrm{diag}(m_u, m_d, m_s), \\
    \tilde{M}_{q} &= 
    \mathrm{diag}\big( m_u (\xi^s_{u} - i \xi^p_{u}), m_d (\xi^s_{d} - i \xi^p_{d}), m_s (\xi^s_{s} - i \xi^p_{s}) \big).
\end{align}
\end{subequations}
Here $m_u$, $m_d$ and $m_s$ denote the up, down and strange quark masses, respectively. 
Inserting Eqs.~\eqref{eq:sigma} and \eqref{eq:field-dep-mass} into the Lagrangian \eqref{eq:lagrangian-chipt} and expanding $\Sigma$, the meson interaction Lagrangian with the dark scalar boson is obtained as 
\begin{align}
\mathcal{L}_{\mathrm{meson}}
&= - \frac{B}{\sqrt{2} v} \pi_a \hat{\phi} \mathrm{Tr} \big( \lambda^a \Im(\tilde{M}_{q})\big) 
    - \frac{B}{2 v f_\pi} \pi_a \pi_b \hat{\phi} \mathrm{Tr} \big( \lambda^a \lambda^b \Re(\tilde{M}_{q}) \big) \nn \\
&\quad 
    + \frac{B}{3 \sqrt{2} v f_\pi^2}\pi_a \pi_b \pi_c \hat{\phi} \mathrm{Tr} \big( \lambda^a \lambda^b \lambda^c \Im(\tilde{M}_{q}) \big) 
%
    + \frac{1}{6 v f_\pi^3} \pi_a \pi_b \pi_c \pi_d \hat{\phi} \mathrm{Tr} \big( \lambda^a \lambda^b \lambda^c \lambda^d \Re(\tilde{M}_{q}) \big) \nn \\
&\quad 
    + \frac{1}{6 f_\pi^3} \pi_a \pi_b \pi_c \pi_d\mathrm{Tr} \big( \lambda^a \lambda^b \lambda^c \lambda^d m_q \big), \label{eq:chiralpt-int}
\end{align}
where the constant term and linear term of $\hat{\phi}$ are omitted. The first term induces the mixing between the dark scalar and light pseudoscalar mesons. It is proportional to $\mathrm{Im}(\tilde{M}_q)$, hence to the pseudoscalar coupling $\xi^p_q$. This mixing with pseudoscalar bosons can enhance the dark scalar decays into di-photons, and three-meson final states via the fifth term. The mass eigenstate of the dark scalar boson $\phi$ can be obtained by diagonalizing the mass matrix of the mesons and dark scalar with this mixing given in Appendix~\ref{apdx:meson-int-masseigenstates}. 
The second and fourth terms induce the dark scalar decays into di-mesons and four-mesons, which are proportional to the scalar coupling $\xi^s_q$. The third term, which is proportional to $\xi^p_q$, induces the decays into three-meson final states. The number of final state mesons reflects the CP structure of the dark scalar interaction. The simultaneous presence of both even and odd numbers of meson interactions is a 
characteristic prediction of the CP violation in our model.

\section{Decay and production of dark scalar boson} \label{sec:decay-dark-scalar}
The dark scalar boson can decay into the charged leptons, photons and the mesons if kinematically allowed. In this section, we present the decay widths of the dark scalar into these final states. We denote $\phi$ as the mass eigenstate of the dark scalar boson in the following discussion, as obtained after solving the mixing with mesons.

\subsection{Decays into charged leptons}
The decay width of $\phi \to \ell \bar{\ell}~(\ell = e, \mu, \tau)$ is given by
\begin{align}
    \Gamma(\phi \to \ell \bar{\ell}) &= \frac{m_\ell^2}{8\pi v^2}  
    \bigg( |\xi^s_{\ell}|^2 \beta_\ell^3 + |\xi^p_{\ell}|^2 \beta_\ell \bigg) m_{\phi}, \label{eq:decay-width-lepton}
\end{align}
where $\beta_\ell = \sqrt{1 - 4m_\ell^2/m_{\phi}^2}$. 

\subsection{Decays into di-photons}
The decay of $\phi \to 2\gamma$ consists of two contributions. One is from the loop contributions 
given in Eq.~\eqref{eq:lag-int}. The decay width from the loop contributions is given by
\begin{align}
    \Gamma(\phi \to 2\gamma) 
    &= \frac{\alpha_{em}^2}{64 \pi^3 v^2} \big(|\xi_{\gamma}|^2 + |\tilde{\xi}_{\gamma} |^2 \big) m_\phi^3.
\end{align}
The other is from the mixing with pseudoscalar mesons which can decay into di-photons. 
The mixing with the pseudoscalar mesons can be taken into account by making the replacement \cite{Domingo:2016yih,Kling:2022uzy}, 
\begin{align}
    \frac{\tilde{\xi}_{\gamma}}{v} \to
    O_{\hat{\phi} \phi} \frac{\tilde{\xi}_{\gamma}}{v}
    + O_{\hat{\phi} \pi^0} C^\gamma [\pi^0] 
    + O_{\hat{\phi} \eta} C^\gamma [\eta] 
    + O_{\hat{\phi} \eta'}C^\gamma[\eta'],
\end{align}
where $O$ is the diagonalization matrix defined in Appendix \ref{apdx:meson-int-masseigenstates}, and $C^\gamma$ are the decay amplitudes of the mesons into $2\gamma$. We use the values of $C^\gamma[\pi^0] = -10.75,~C^\gamma[\eta] = -10.8,~C^\gamma[\eta'] = -13.6$ in GeV$^{-1}$ given in \cite{Domingo:2016yih}. In our analysis, we numerically diagonalize the mass matrix of the mesons and dark scalar to obtain the mixing matrix $O$.

Including the mixing, the decay width into $2\gamma$ is obtained as 
\begin{align}
    \Gamma(\phi \to 2\gamma) 
    &= \frac{\alpha_{em}^2}{64 \pi^3}  m_\phi^3 \bigg(\bigg|\frac{\xi_{\gamma}}{v}\bigg|^2  \nn \\
    &\quad
    + \bigg| O_{\hat{\phi} \phi} \frac{\tilde{\xi}_{\gamma}}{v} 
    + O_{\hat{\phi} \pi^0} C^\gamma [\pi^0] 
    + O_{\hat{\phi} \eta} C^\gamma [\eta] 
    + O_{\hat{\phi} \eta'}C^\gamma[\eta']\bigg|^2 \bigg). \label{eq:decay-width-2gamma}
\end{align}

\subsection{Hadronic decays into two and four-meson final states}
For $m_\phi \leq 2$ GeV, the meson Lagrangian should be used to express the decays into mesons. The decay widths into two pions and two kaons are given by \cite{Donoghue:1990xh}
\begin{align}
\Gamma_{\pi \pi} &= \frac{3 G_{F}}{16 \sqrt{2} \pi m_\phi} \beta_{\pi}\left|
\xi_g \frac{2}{27}(\Theta_{\pi}-\Gamma_{\pi}-\Delta_{\pi})+
\frac{m_u \xi_u^s +m_d \xi_d^s} {m_u+m_d} \Gamma_{\pi} + \xi_s^s\Delta_{\pi}
\right|^{2}, \\
\Gamma_{K K} &= \frac{ G_{F}}{4 \sqrt{2} \pi m_\phi} \beta_{K}\left|
\xi_g \frac{2}{27}(\Theta_{K}-\Gamma_{K}-\Delta_{K})+
\frac{m_u \xi_u^s + m_d \xi_d^s}{m_u+m_d} \Gamma_{K} + \xi_s^s\Delta_{K}
\right|^{2},
\label{eq:2K-width}
\end{align}
where $\Theta_{\pi,K}$, $\Gamma_{\pi,K}$ and $\Delta_{\pi,K}$ are form factors evaluated at $\sqrt{s}=m_{\phi}$. In our analysis, we use the form factors extracted through dispersion relations~\cite{Winkler:2018qyg, Monin:2018lee} to take into account the higher order effects.   

For the decay width into other mesons $\eta\eta,~\rho\rho$ including 4 body decays $\pi\pi\pi\pi,~\pi\pi K\bar{K},~\cdots$, we use the following expression instead of 
the coupling in Eq.~\eqref{eq:chiralpt-int} \cite{Grinstein:1988yu,Winkler:2018qyg}
\begin{align}
    \Gamma_{\mathrm{even~mesons}} &= C_{\mathrm{mesons}} |\xi_g|^2 m_\phi^3 \beta_{2\pi},
\end{align}
where $C_{\mathrm{mesons}}=5.1 \times 10^{-9}$ GeV$^{-2}$ to obtain a smooth hadronic decay rate transiting into the rate of the spectator model at $m_\phi = 2$ GeV.

\subsection{Hadronic decays into three-meson final states}
The dark scalar can also decay into three light mesons such as $\pi\pi\pi,~\eta \pi\pi,~\eta' \pi\pi,~\pi K\bar{K}$. The partial widths of $\phi \to \Pi_i\Pi_j\Pi_k$ are given by \cite{Domingo:2016yih,Kling:2022uzy}\footnote{
The decay widths into three mesons may change in the presence of kinetic mixing between the CP-odd scalar and pseudoscalar mesons \cite{DallaValleGarcia:2023xhh,Ovchynnikov:2025gpx,Kryshtal:2026bhq}.}

\begin{align}
\Gamma(\phi \to \Pi_i\Pi_j\Pi_k)&=\frac{1}{256S_{ijk}\pi^3m_\phi}\int_{(m_j+m_k)^2}^{(m_\phi-m_i)^2}ds  |\mathcal{M}_\phi^{ijk}|^2\\ 
&\quad \times \sqrt{1-\frac{2(m_j^2+m_k^2)}{s}+\frac{(m_j^2-m_k^2)^2}{s^2}} \sqrt{\biggl(1+\frac{s-m_i^2}{m_\phi^2}\biggr)^2-\frac{4s}{m_\phi^2}},
\label{eq:trimeson-width}
\end{align}
where $S_{ijk}$ is a symmetry factor, $1!$, $2!$, $3!$, depending on the number of identical particles in the final state. The decay amplitude $\mathcal{M}_\phi^{ijk}$ is given by
\begin{align}
\mathcal{M}_\phi^{ijk} = O_{\hat{\phi} \phi}\mathcal{A}_\phi^{ijk} + \sum_l O_{\hat{\phi} l}\mathcal{A}^{ijkl}.
\end{align}
Note that $\mathcal{A}_\phi^{ijk}$ is proportional to the pseudoscalar couplings.
For $m_\phi \leq 1.3$ GeV, concrete expressions for $\mathcal{A}_\phi^{ijk}$ and $\mathcal{A}^{ijkl}$ can be read from the Lagrangian Eq.~\eqref{eq:chiralpt-int} and are  presented in Appendix \ref{apdx:meson-couplings}. On the other hand, for $1.3~\mathrm{GeV} \leq m_\phi \leq 3$ GeV, the spectator model is used to calculate the amplitudes \cite{Dolan:2014ska,Winkler:2018qyg}.

\subsection{Decays into quarks}
The dark scalar can decay into quarks through both $\xi^s_q$ and $\xi^p_q$. Depending on the mass range discussed in the hadronic decays, the decay widths can be divided as
\begin{align}
    \Gamma(\phi \to q\bar{q}) &= 
    \begin{cases}
        \frac{m_q^2}{8\pi v^2} |\xi^s_{q}|^2 \beta_q^3 m_{\phi},~~&(2~\mathrm{GeV} < m_\phi < 3~\mathrm{GeV}) \\
    \frac{m_q^2}{8\pi v^2}  
    \big( |\xi^s_{q}|^2 \beta_q^3 + |\xi^p_{q}|^2 \beta_q \big) m_{\phi},~~&( m_\phi > 3~\mathrm{GeV} )
    \end{cases}.
\end{align}

\subsection{Decay branching ratio and decay length}
\begin{figure}[t]
\centering
\includegraphics[scale=0.4]{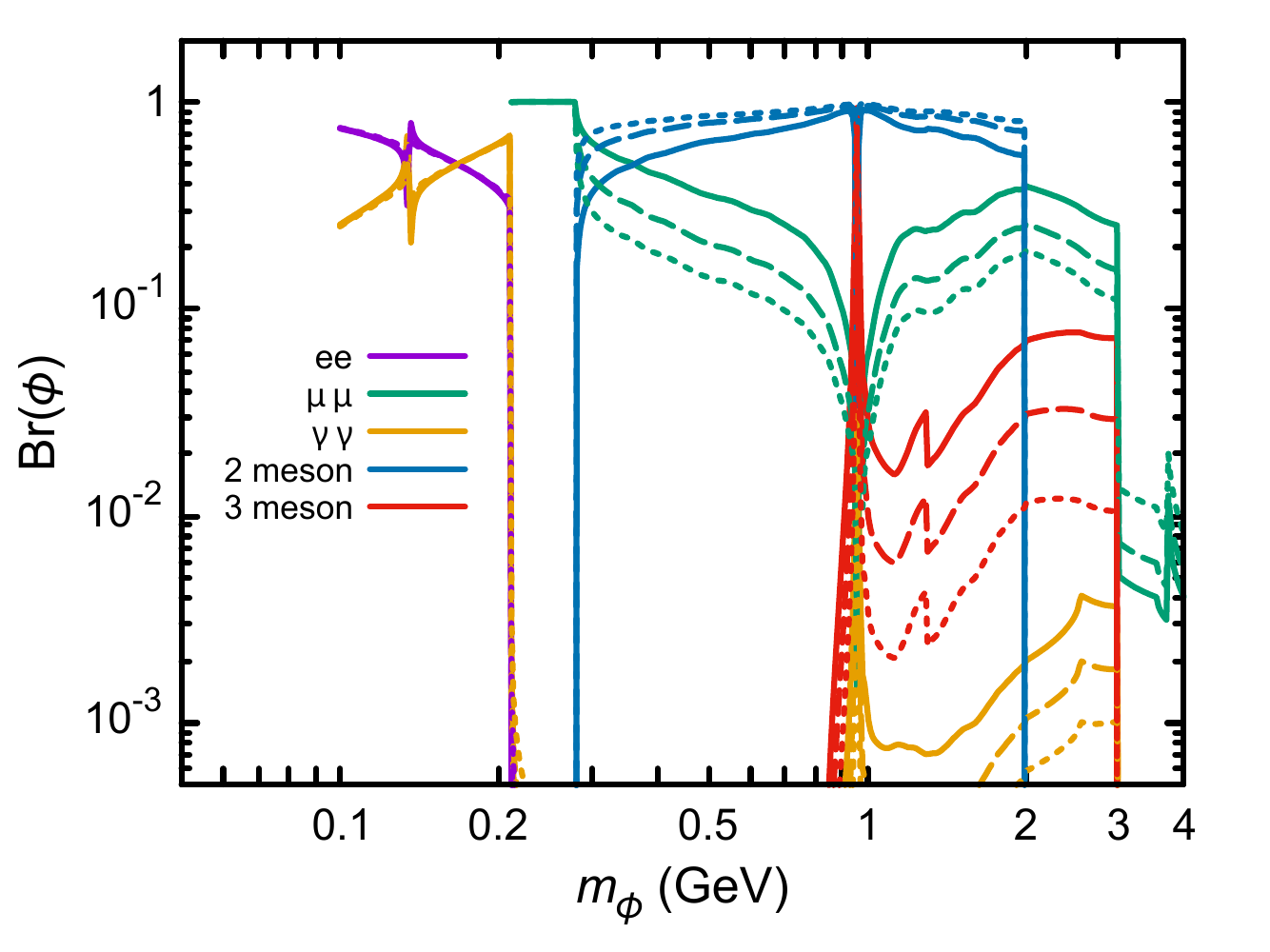} 
\hspace*{-1cm}
\begin{tabular}{cc}
    \includegraphics[scale=0.4]{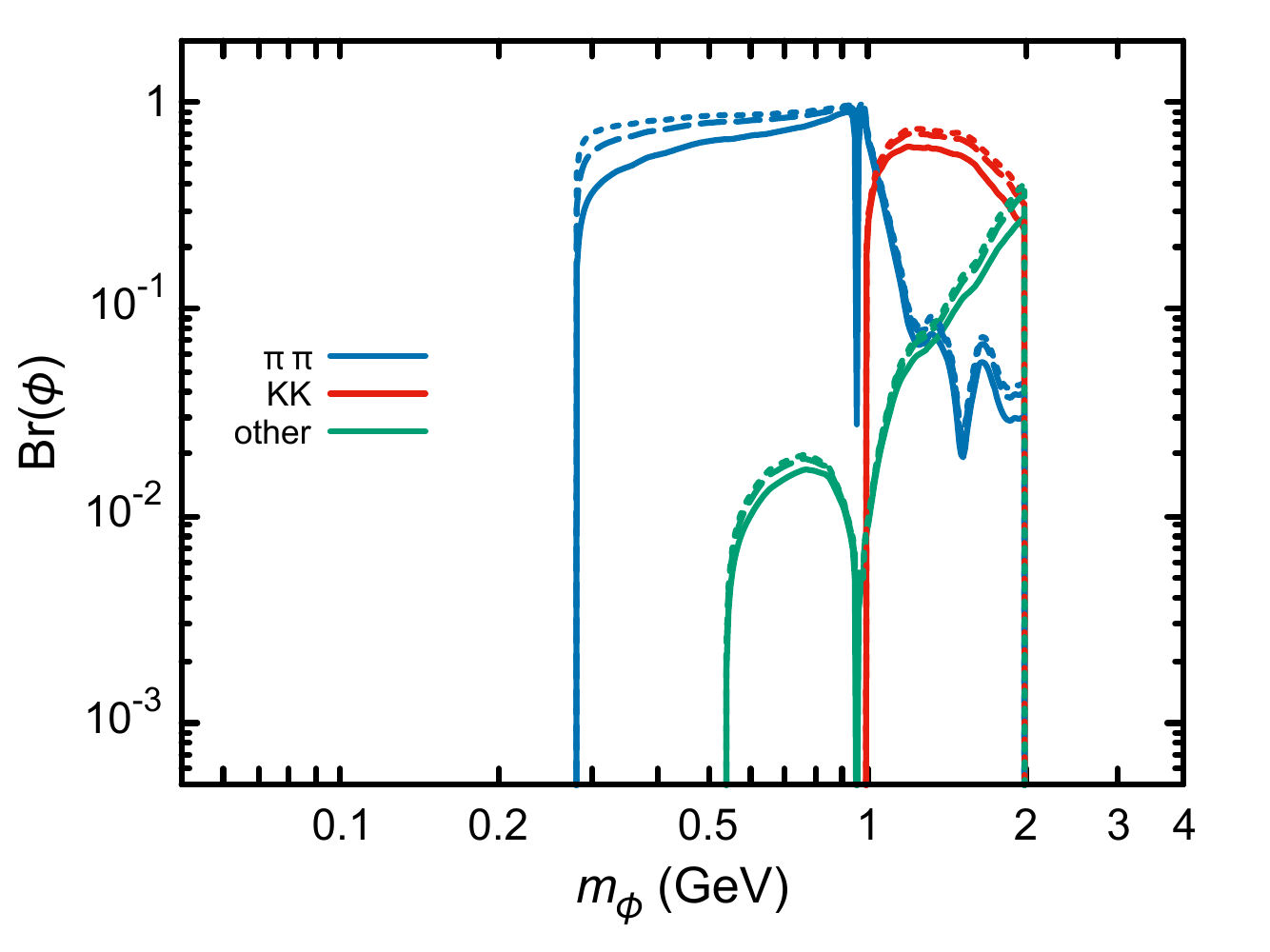} &
    \includegraphics[scale=0.4]{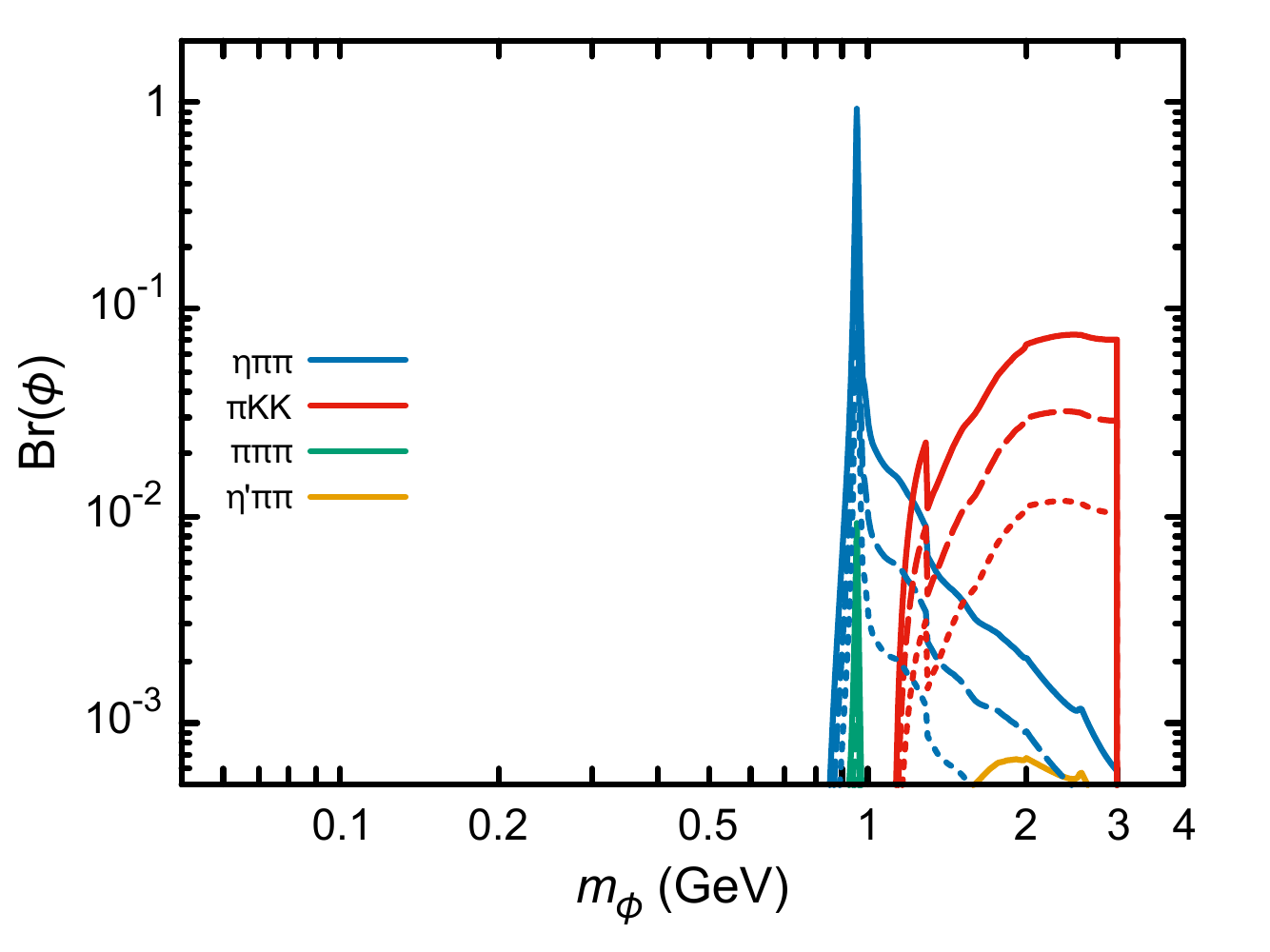} 
\end{tabular}
    \caption{
    The branching ratios of the dark scalar decays (Top), and those into two-meson final states (Left Bottom) and into three-meson final states (Right Bottom). The parameter $\theta$ is varied as $\theta=\pi/3$ (solid), $\pi/4$ (dashed) and $\pi/6$ (dotted), and $\epsilon$ is fixed to $10^{-5}$.}
\label{fig:branching-ratio}
\end{figure}
We show the decay branching ratios and decay length of the dark scalar boson. For convenience, we parameterize the scalar and pseudoscalar coupling as
\begin{align}
    \xi^s_f = \epsilon \cos\theta,\quad
    \xi^p_f = -\epsilon \sin\theta, \label{eq:eps-theta}
\end{align} 
where $\theta$ determines the magnitude of CP violation, and $\theta = 0$ and $\pi/2$ correspond to a pure CP-even and an odd scalar, respectively. The overall magnitude of the couplings is controlled by $\epsilon$, which is assumed to be universal for all fermions.

We show the branching ratios of the dark scalar in Figure \ref{fig:branching-ratio}. The top panel is the decay branching ratios of the dark scalar boson into electrons, muons, photons as well as two mesons and three mesons. For illustration purposes, we take $\epsilon = 10^{-5}$. Solid, dashed and dotted curves correspond to $\theta=\pi/3,~\pi/4$ and $\pi/6$, respectively. For $2 m_{\pi^0} \leq m_\phi < 2 m_\eta$, the decays into two mesons are dominant. The branching ratios into three-meson final states are one or two orders of magnitude below that of the two-meson decays for $2m_{K} \leq m_\phi < 2 m_\eta$. Above the eta pair threshold, two-meson decays are much suppressed because the dark scalar can decay into quarks. Thus, the CP violation in the dark scalar decay can be examined between $1~\mathrm{GeV} \lesssim m_\phi \lesssim 2$ GeV in our approach.  
The bottom left and right panels show the contributions from individual decay modes for two-meson and three-meson final states. One can see that the comparisons between the $KK$ final states and $\pi KK$ are good candidates for probing CP violation. The muon pair final states are also available as we will show later.

\begin{figure}[t]
\centering
    \includegraphics[scale=0.4]{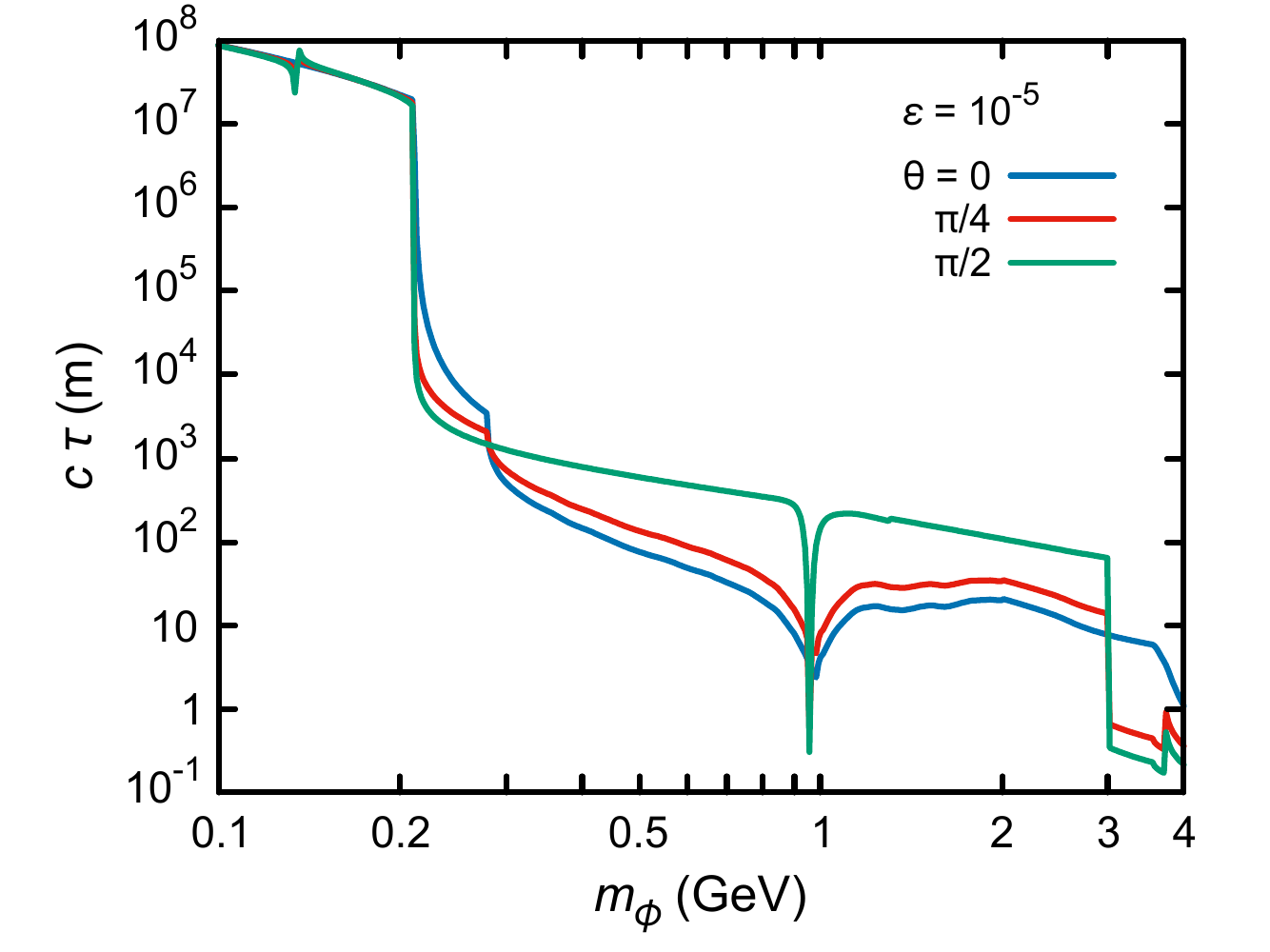} 
    \caption{Decay length of CP-violating dark scalar boson. The blue, red and green curves correspond to $\theta=0,\pi/4$ and $\pi/2$, respectively. The parameter $\epsilon$ is fixed to $10^{-5}$.
    }
\label{fig:decay-length}
\end{figure}
The decay length of the dark scalar at rest, $c\tau_\phi$, is shown in Figure \ref{fig:decay-length} for $\theta=0,~\pi/4$ and $\pi/2$ with $\epsilon=10^{-5}$. The sharp dip at $m_\phi\simeq1~\mathrm{GeV}$ originates from the $\eta'$ mixing and the nearby $K \bar{K}$ threshold.
The decay length changes from $\mathcal{O}(10)$\,m to $\mathcal{O}(10^3)$\,m depending on $\theta$ above $m_\phi > 1$ GeV. 
In the SHiP and FASER2 experiments, the decay volumes are located at distances of $33$\,m and $650$\,m from their respective beam collision points. Taking the Lorentz boost into account, the shorter decay length than these distances would be favored. Thus, $\theta$ determines not only the branching ratio but also the decay length which results in the dependence of the decay probability on $\theta$.

\subsection{Production of dark scalar boson} 
The dark scalar boson can be mainly produced through decays of heavy mesons such as $B,~K$ and $\eta$. However, since the dark scalar mass in our interest is above $1$ GeV, it is enough to consider the $B$ meson decays as a dark scalar production process. 

The decay branching ratio of the $B$ meson into the dark scalar and a strange meson is 
given by \cite{Dolan:2014ska,Kling:2022uzy,DallaValleGarcia:2023xhh,Ovchynnikov:2025gpx},
\begin{align}
    \mathrm{Br}(B \to X_s + \phi) = \big( 5.7 |\xi_t^s|^2 + 26.2 |\xi^p|^2 \big) \left(1 - \frac{m_\phi^2}{m_b^2}\right)^2, \label{eq:br-B}
\end{align}
where $\xi^s_t$ is the modifier for the top scalar-interaction and $m_b$ is the bottom quark mass, $m_b=4.18$ GeV. The coefficient $\xi^p$ is 
\begin{align}
\xi^p &= \xi^p_t + \xi^p_b - 0.01 \xi^p_s,
\end{align}
where $\xi^p_{t,b,s}$ are the modifiers for the pseudoscalar interaction of top, bottom and strange quarks, respectively. 
The first and second terms in Eq.~\eqref{eq:br-B} represent the contributions from the CP-even and odd components of the dark scalar. For the universal modifiers given in Eq.~\eqref{eq:eps-theta}, the CP-odd contribution is about $18$ times larger than the CP-even one,
\begin{align}
    \mathrm{Br}(B \to X_s + \phi) = \big( 5.7 \cos^2\theta + 105 \sin^2\theta \big) \epsilon^2 \left(1 - \frac{m_\phi^2}{m_b^2}\right)^2. \label{eq:br-B-common}
\end{align}

\begin{figure}[t]
\centering
    \includegraphics[scale=0.4]{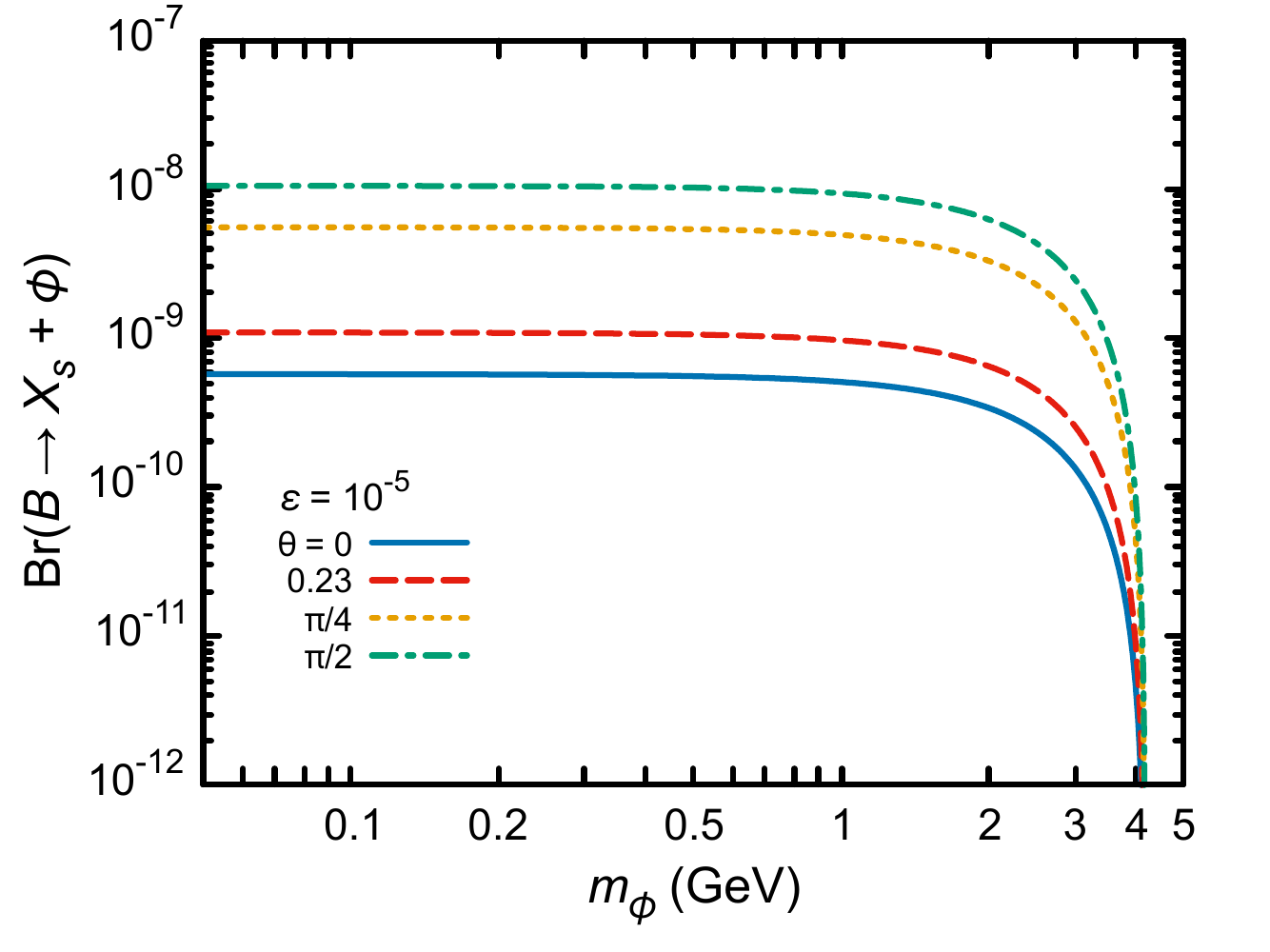}
    \caption{
    Branching ratio of $B$ meson decays into CP-violating dark scalar boson. 
    }
\label{fig:meson-br}
\end{figure}
The meson decay branching ratio is shown in Figure \ref{fig:meson-br}. The parameters are indicated in the figure. The CP-even and odd contributions become comparable for $\theta=0.23 \simeq \pi/13.6$. As $\theta$ becomes larger, the dark scalar is produced more from the $B$ meson decay.

\section{Numerical Results} \label{sec:results}
\subsection{Expected number of Signal Events}
The dark scalar boson can decay into photons, muons, pions, kaons, eta mesons, and eta-prime mesons in the mass range of our interest. The charged pion and kaon are sufficiently long-lived so that these particles penetrate the SHiP and FASER2 detectors while leaving tracks. On the other hand, the neutral mesons, $\pi^0$ and $K_S$ as well as $\eta$ and $\eta'$, mainly decay into photons and pions. Since both experiments can identify charged particles and photons \cite{SHiP:2021nfo,SHiP:2025ows,FPF:2025bor},  these dark scalar decays will be used as signal.  The other neutral meson, $K_L$, will escape the detector before it decays. When we approximate the neutral and charged kaon masses as being the same, the branching ratios into each kaon become equal. Then, we will subtract the $K_L$ final state by multiplying the decay branching ratio of $\phi \to KK$ in Eq.~\eqref{eq:2K-width} by $3/4$. 
For three-meson final states, the $K_L$ contribution is removed channel by channel. The $\pi^0 K^+ K^-$ mode is unchanged, whereas the $\pi^0 K^0\bar K^0$, $\pi^+ K^- K^0$, and $\pi^- K^+ \bar K^0$ modes are each multiplied by $1/2$ in the neutral-kaon mass-degenerate approximation.

Based on the above discussion, we define the dark scalar decays into two-meson and three-meson final states, as well as charged lepton pairs and di-photon final states, as the signal events. The expected number $N_i$ of specific signal events $i$ is given by
\begin{align}
   N_i &= N_{\rm pot} \int dp_B d\theta_B~f_{B}(p_B,\theta_B)
   {\rm Br}(B \to X_s + \phi)~
   {\rm Br}(\phi \to i)~
   \mathcal{P}_{\phi}^{\rm det}~, \label{eq:expected-num-of-events}
\end{align}
where $N_{\rm pot}$ is the number of protons on target and $f_{B}(p_B,\theta_B)$ is the distribution function of the $B$ meson. The momentum of $B$ mesons is denoted as $p_B$, and the polar angles of $p_B$ with respect to the beam direction are defined by $\theta_B$. For the FASER2 experiment, $N_{\rm pot}$ and  $f_{B}(p_B,\theta_B)$ should be replaced with the integrated luminosity $\mathcal{L}$ and the differential cross section with respect to $p_B$ and $\theta_B$, respectively.
The branching ratio of the $B$ meson is given by Eq.~\eqref{eq:br-B} and those of $\phi$ are given in Section~\ref{sec:decay-dark-scalar}.

\begin{table}[t]
\centering
\begin{tabular}{|c|c|c|c|c|c|c|c|} \hline
 \hspace{2cm} & ~~$L_{\rm min}$~~ & ~~$L_{\rm max}$~~ & ~~$W_{\mathrm{min}}$~~ & ~~$W_{\mathrm{max}}$~~ & ~~$H_{\mathrm{min}}$~~ & ~~$H_{\mathrm{max}}$~~ & ~~$N_{\mathrm{pot}}$/$\mathcal{L}$~~ \\ \hline 
  SHiP     & 33.5 & 83.5 & 1.0 & 4.0 & 2.7 & 6.0 & $6 \times 10^{20}$ \\ \hline
 FASER2  & 650 & 660 & 2.6 & 2.6 & 1.0 & 1.0 & 3\,ab$^{-1}$ \\ \hline
\end{tabular}
\caption{
Dimensions of the SHiP~\cite{SHiP:2015vad,SHiP:2020vbd,ShipECN3} and FASER2~\cite{Adhikary:2024nlv,Feng:2025adw,FPF:2025bor} detectors and the number of protons on target ($N_{\mathrm{pot}}$)/the integrated luminosity ($\mathcal{L}$) used in this study. 
$L_{\rm min}$ and $L_{\rm max}$ are the distances to the front and rear end of the detector from the interaction point, respectively. $W_{\mathrm{min/max}}$ and  $H_{\mathrm{min/max}}$ are the width and height of the rectangular detectors for SHiP and FASER2 at the front/rear face, respectively. The unit of the distance and length is in meters.
}
\label{tab:detector-dimension}
\end{table}
The probability that the dark scalar decays inside the detector is given in \cite{Araki:2026wfe} by
\footnote{The decay probability may change with realistic detector geometry and acceptance for daughter particles. Such a setup can be implemented using  \cite{Ovchynnikov:2023cry}.}
\begin{align}
 \mathcal{P}_{\phi}^{\rm det} =
 \left( 
    e^{-\frac{L_{\mathrm{min}}}{\beta\gamma d_\phi}} - e^{-\frac{L_{\mathrm{max}}}{\beta\gamma d_\phi}}
 \right) 
 \Theta_x 
 \left( 
    \frac{H_{\rm max}}{2} - L_\mathrm{max}~ \left| \frac{p_{\phi, x}}{p_{\phi, z}} \right|
 \right)
 \Theta_y 
 \left( 
    \frac{W_{\rm max}}{2} - L_\mathrm{max}~ \left|\frac{p_{\phi, y}}{p_{\phi, z}} \right|
 \right)~.
 \label{eq:P_phi_rect}
\end{align}
The decay length of the dark scalar is given by $d_\phi (= c \tau_\phi)$ times the Lorentz factor $\beta\gamma$.
The step functions $\Theta_x$ and $\Theta_y$ are multiplied to restrict the momentum of a dark scalar, $(p_{\phi, x}$, $p_{\phi, y}$, $p_{\phi, z})$, in order for the dark scalar to pass through the detector. The distances, heights and widths of the detector are given in Table~\ref{tab:detector-dimension}.

\subsection{Constraints} 
Light CP-even and CP-odd dark scalars have been searched for independently in various experiments, e.g., collider experiments, beam dump experiments, and rare meson decay searches (see review \cite{D_brich_2019,Kling:2022uzy,Ferber:2023iso}). For a CP-even dark scalar with a few GeV mass, the most stringent upper bound on the scalar mixing has been set by the LHCb experiment \cite{LHCb:2015nkv,LHCb:2016awg,Winkler:2018qyg}. Reinterpreting the bound, $\epsilon$ is constrained to be $\lesssim 10^{-4}$ for $m_\phi \simeq 2$ GeV. For a  CP-odd dark scalar, the most stringent upper bound on the fermion couplings has been set by the CHARM experiment \cite{CHARM:1985anb,CHARM:1985nku,D_brich_2019}. The result gives the upper bound $\epsilon \lesssim 10^{-5}$ for the same $m_\phi$. 

On the other hand, the non-vanishing CP-violating phase $\theta$ can be constrained by EDM measurements. The current upper bound on the electron EDM is $|d_e| < 4.1 \times 10^{-30}~e\cdot\mathrm{cm}$ \cite{Roussy:2022cmp,Caldwell:2022xwj}. The dark scalar boson with the effective interaction Eq.~\eqref{eq:lag-int} can contribute to the electron EDM via the Barr-Zee diagram \cite{Barr:1990vd}. The contribution is given by \cite{Brod_2013}
\begin{align}
\frac{d_e^{\rm BZ}}{e} = \frac{16}{3} \frac{\alpha_{\rm em}}{(4\pi)^3} \frac{m_e}{v^2}
\left[ \xi_e^s \xi_t^p f_1 \left(\frac{m_t^2}{m_\phi^2}\right)
+ \xi_e^p \xi_t^s f_2 \left(\frac{m_t^2}{m_\phi^2}\right)
\right],
\end{align}
where $f_1(x)$ and $f_2(x)$ are the loop functions \cite{Brod_2013}. Using the parametrization in Eq.~\eqref{eq:eps-theta}, the electron EDM is given by
\begin{align}
|\epsilon^2 \sin2\theta | \lesssim 1.2 \times 10^{-4},
\end{align}
for the dark scalar mass $1~\mathrm{GeV} \leq m_\phi \leq 2$ GeV. Thus, even for $\theta=\pi/4$, $\epsilon$ can be constrained to be $\lesssim 0.01$. This is much weaker than those from the LHCb and CHARM experiments. In our numerical analysis, we will take $\epsilon = 10^{-5}$ and $5 \times 10^{-6}$ as benchmark points, which evade all the constraints explained above.

\subsection{Numerical Results} 
We present numerical results of the expected number of signal events defined in Eq.~\eqref{eq:expected-num-of-events} with the parametrization in Eq.~\eqref{eq:eps-theta} at the SHiP and FASER2 experiments. 
The detector dimensions and the number of protons on target or integrated luminosity are given in Table~\ref{tab:detector-dimension}. 
Following \cite{Araki:2026wfe}, we apply a momentum cut to the dark scalar: $p>40$ GeV and $100$ GeV for the SHiP and FASER2 experiments in order to avoid unexpected backgrounds, respectively.
In the following plots, we show the regions where at least three signal events are expected from the two- and three-meson decay modes. For comparison, we also show the corresponding regions for the two-muon and di-photon decay modes.
%

\subsubsection{SHiP}

\begin{figure}[t]
\hspace*{-1.0cm}
\centering
\begin{tabular}{cc}
    \includegraphics[scale=0.4]{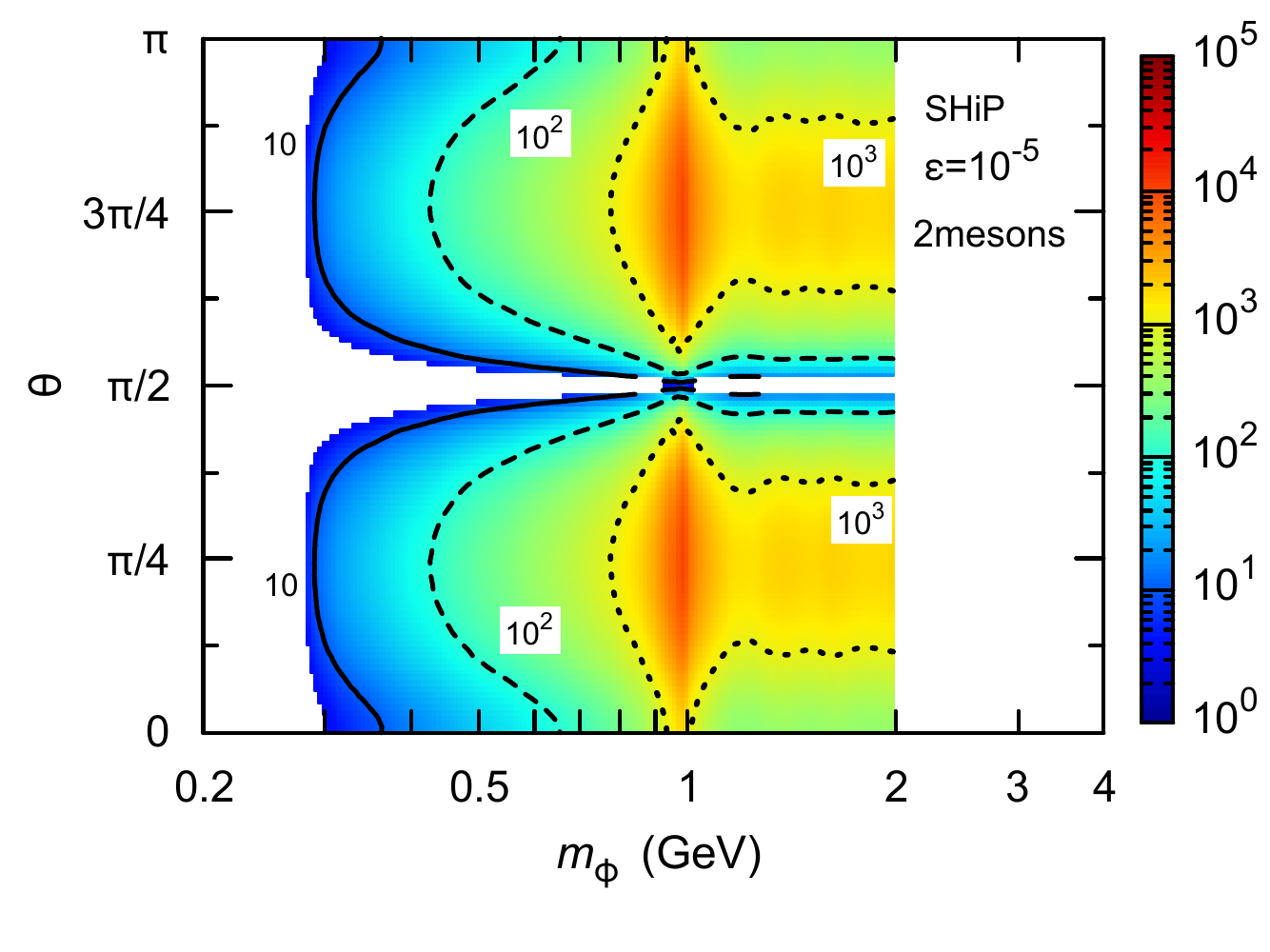} &
    \includegraphics[scale=0.4]{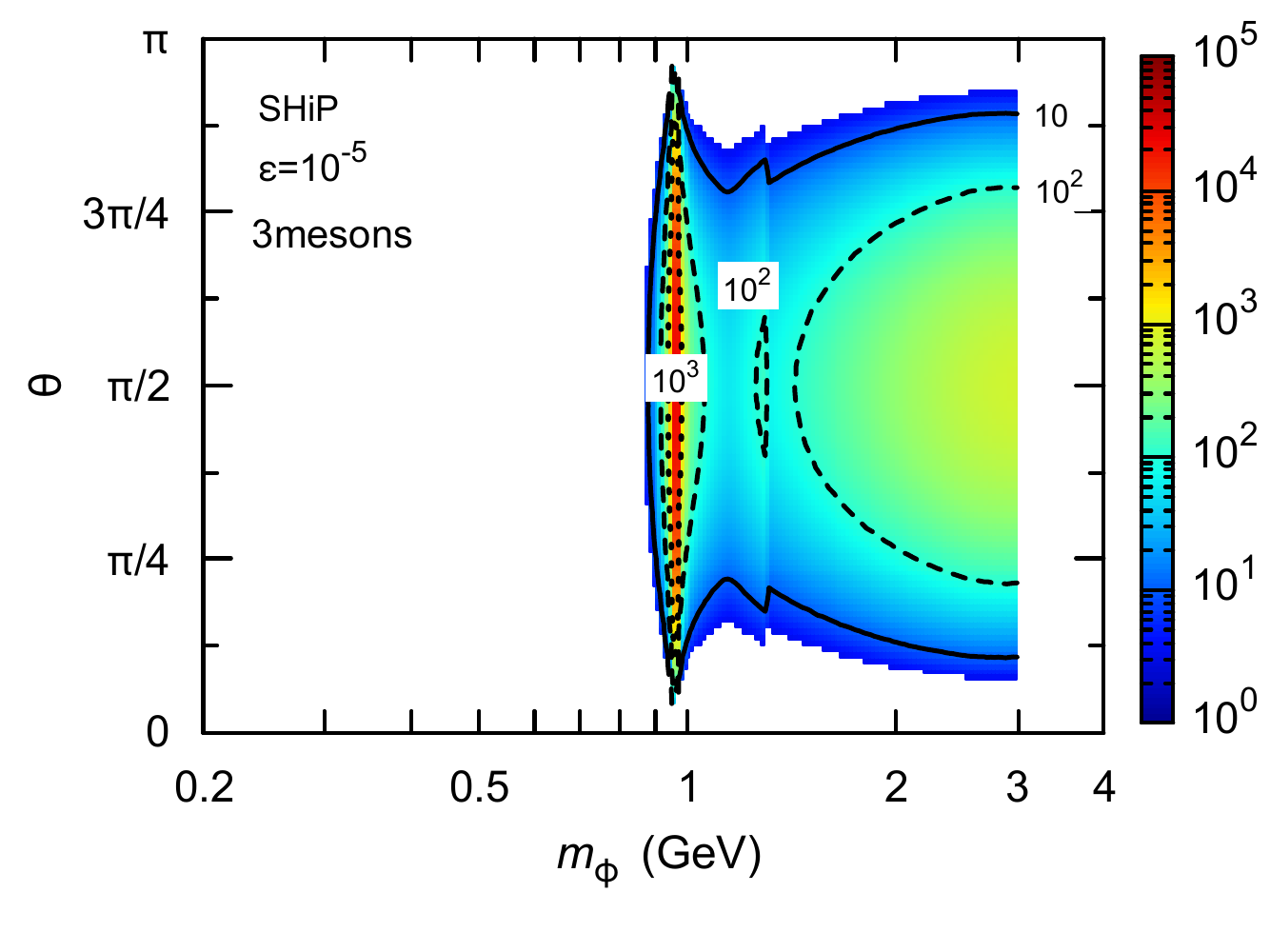} 
\end{tabular}
    \caption{Expected number of the dark scalar decays into predominantly two-meson (Left) and three-meson (Right) final states in the $(m_\phi, \theta)$ plane. Color indicates the expected number of the decays shown in the color bar. The contours correspond to the expected numbers of decays indicated next to each curve. The parameter $\epsilon$ is fixed to $10^{-5}$.}
\label{fig:num-of-mesons-epsf1e-5-ship}
\end{figure}
Figure \ref{fig:num-of-mesons-epsf1e-5-ship} shows the expected numbers of the dark scalar boson decays into predominantly two-meson (left panel) and three-meson (right panel) final states in the $(m_\phi, \theta)$ plane as a color map. We set $\epsilon = 10^{-5}$ in these figures as explained above. The contours correspond to the expected numbers of decays indicated next to each curve. The two-meson final states include the $\pi\pi$, $KK$, and $\eta\eta$ final states, but also include the even-meson final states $\pi\pi\pi\pi$ and $\pi\pi KK$. 
%
\begin{figure}[t]
\hspace*{-1.0cm}
\centering
\begin{tabular}{cc}
    \includegraphics[scale=0.38]{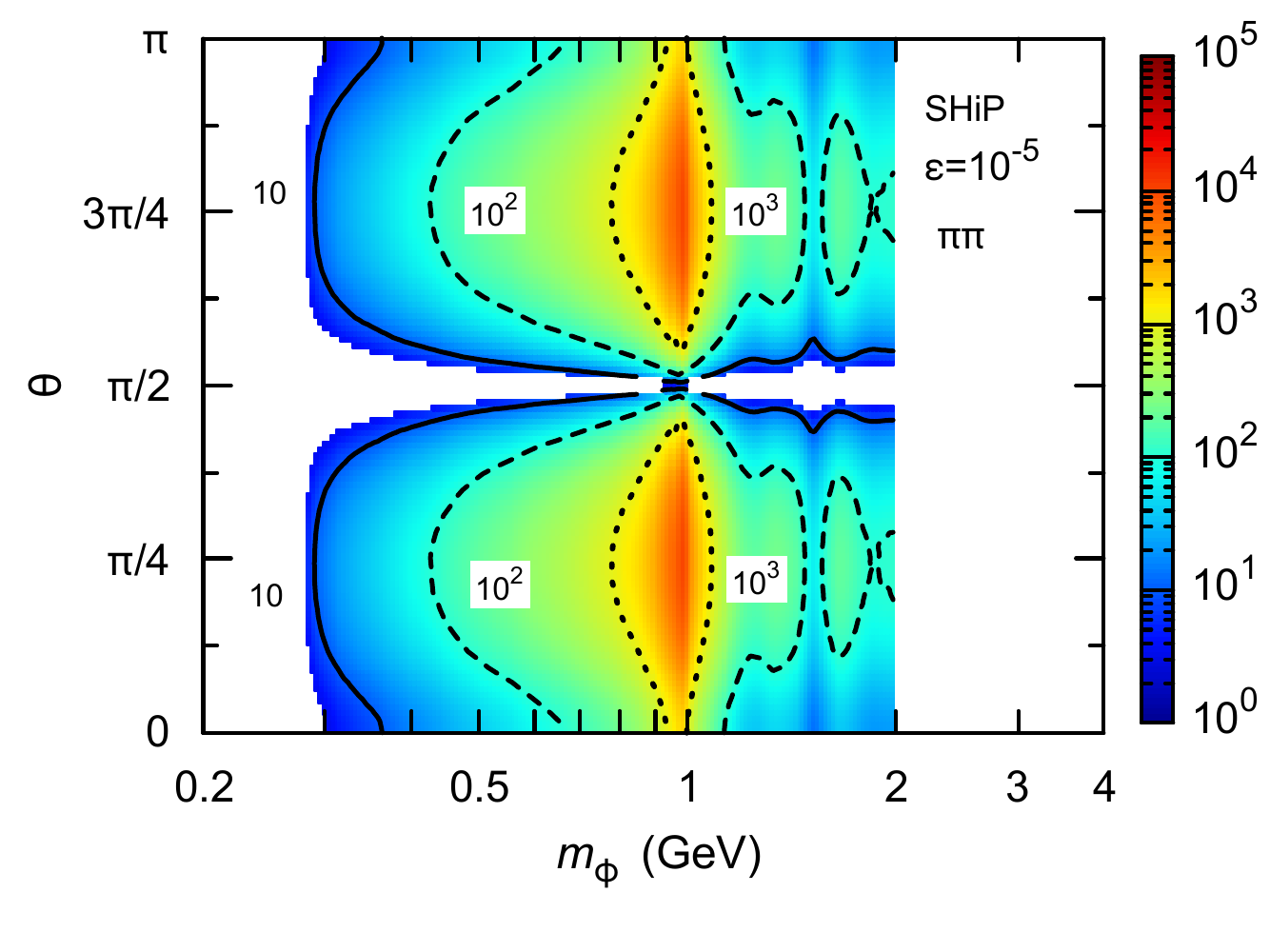} &
    \includegraphics[scale=0.38]{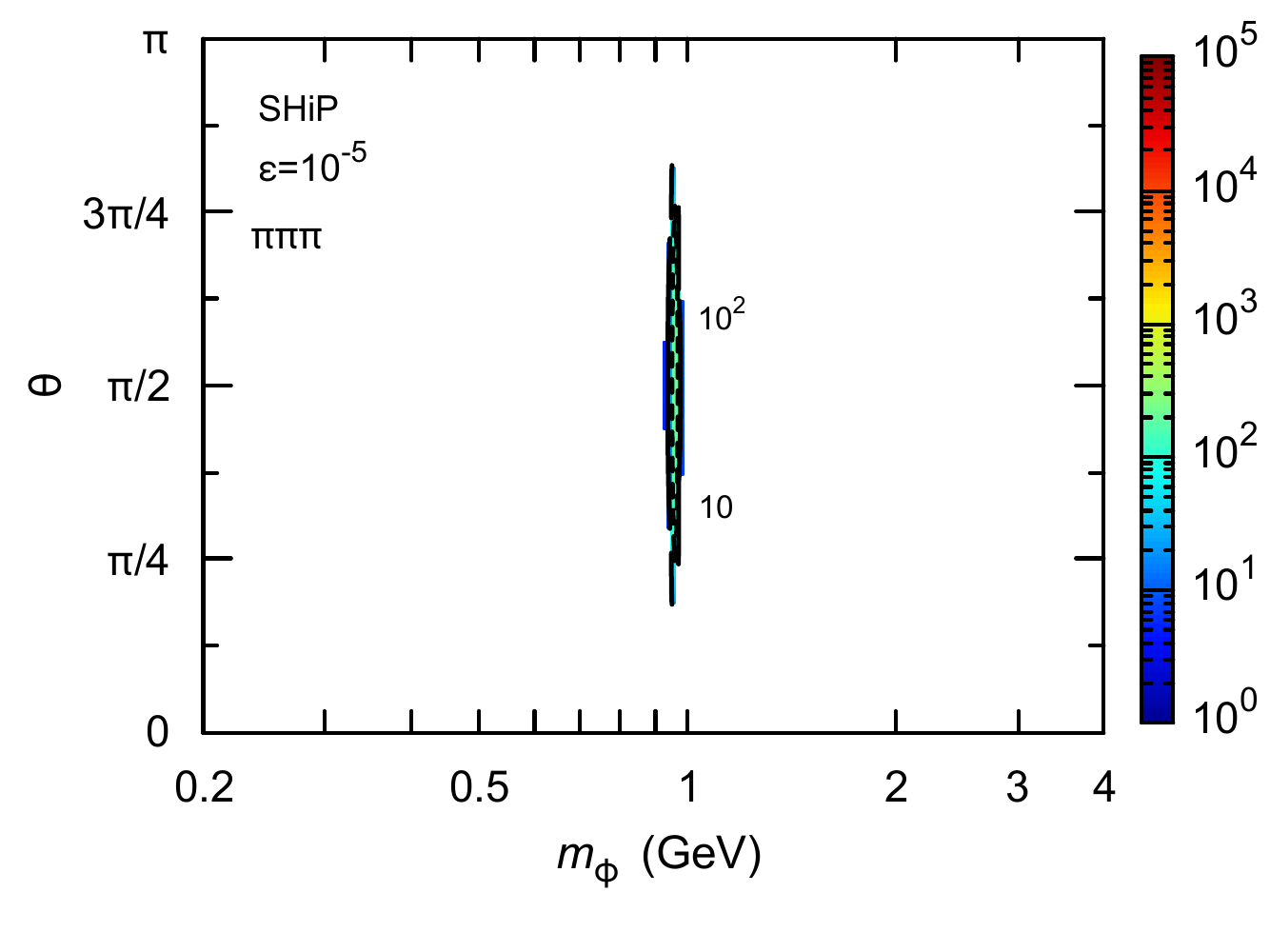} \\
    \includegraphics[scale=0.38]{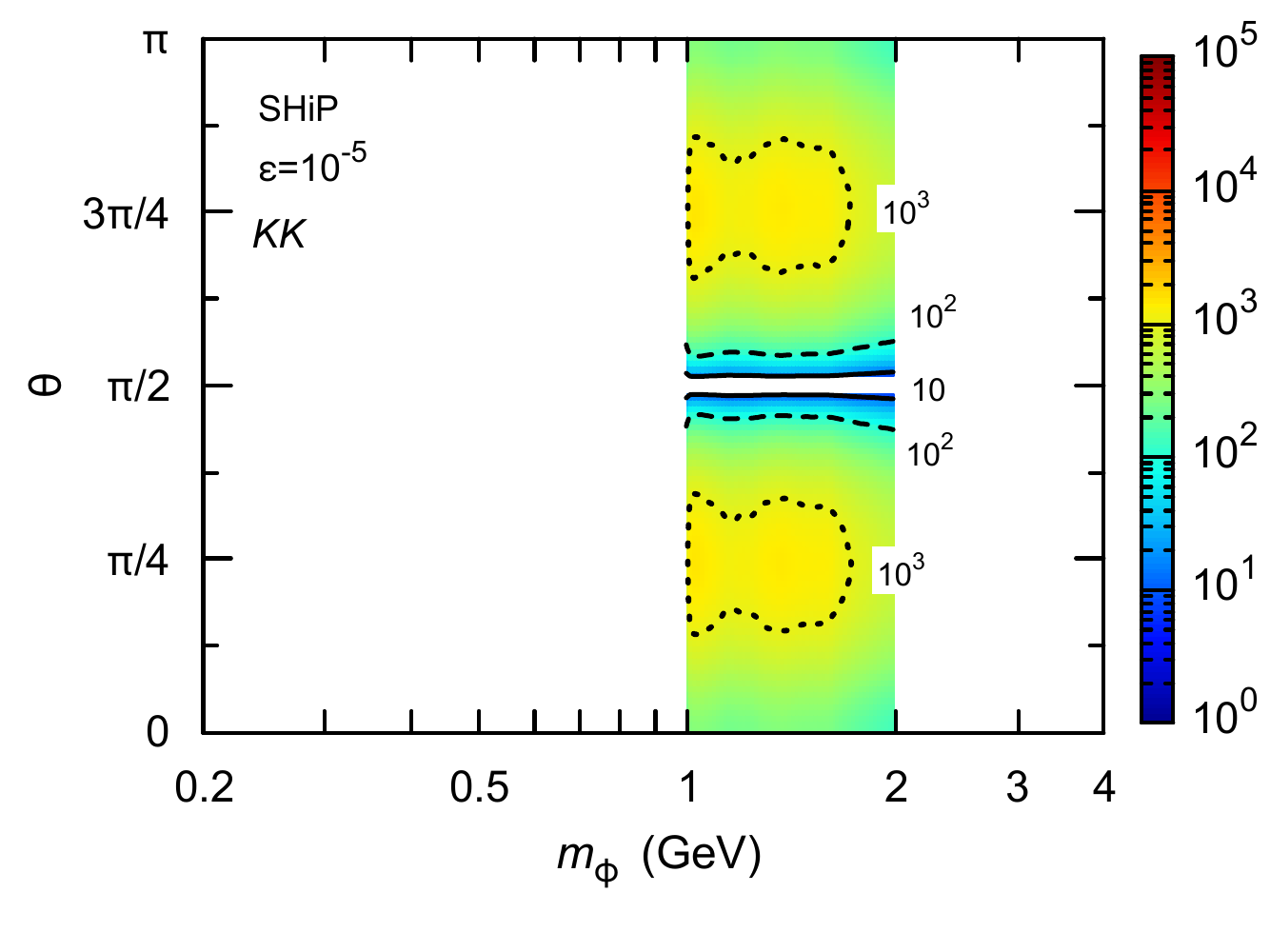} &
    \includegraphics[scale=0.38]{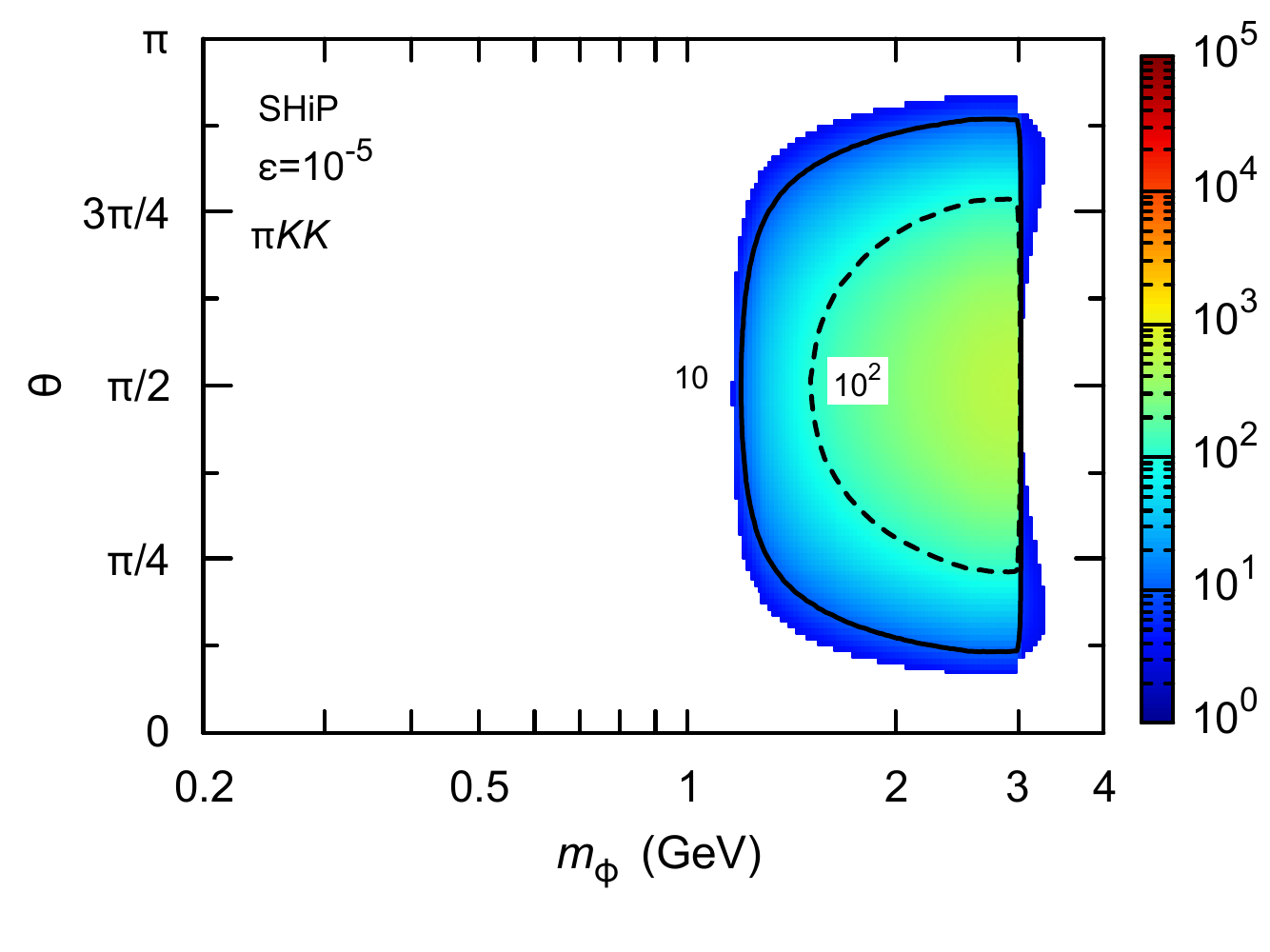}  \\
    \includegraphics[scale=0.38]{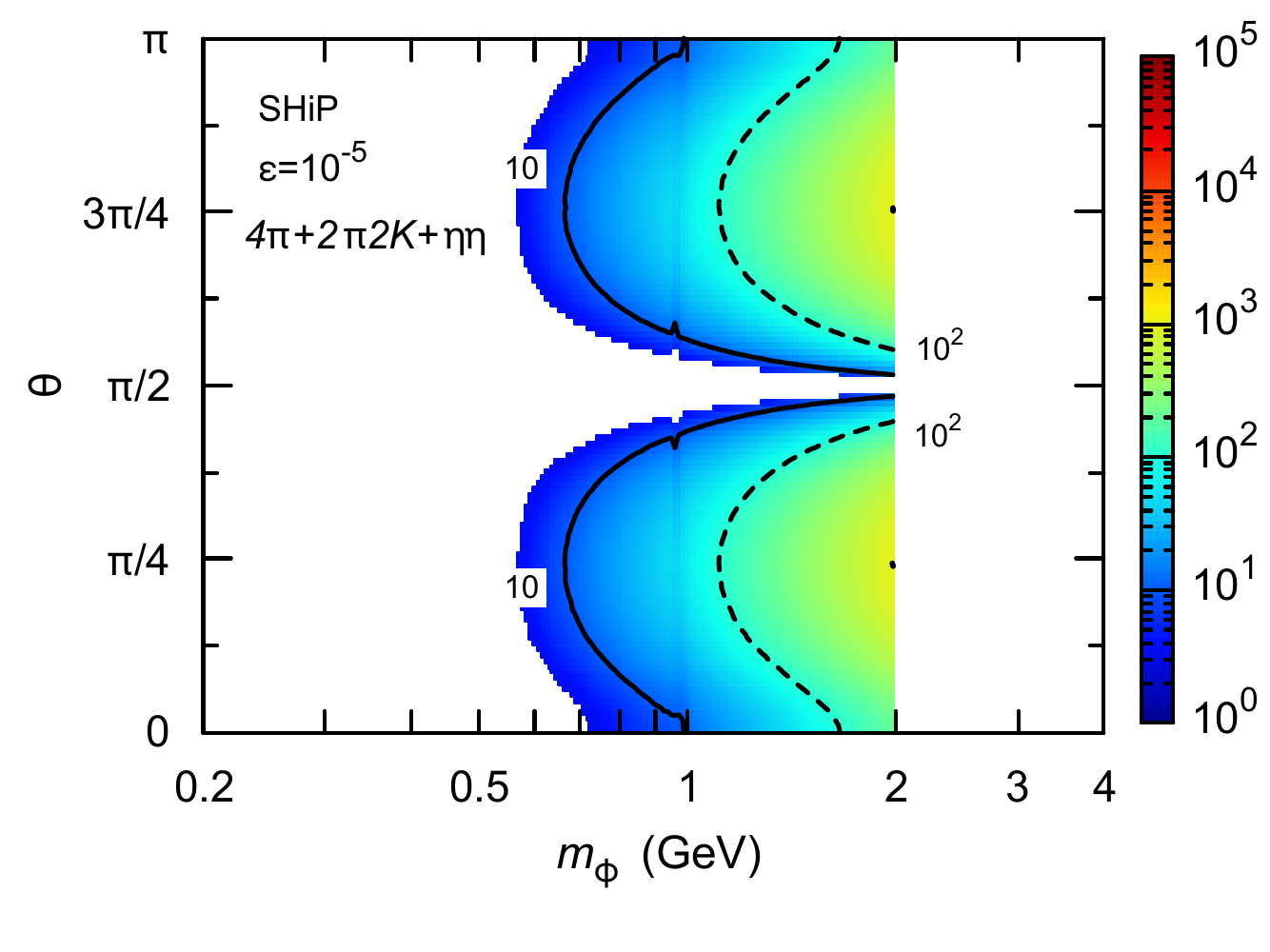} &
    \includegraphics[scale=0.38]{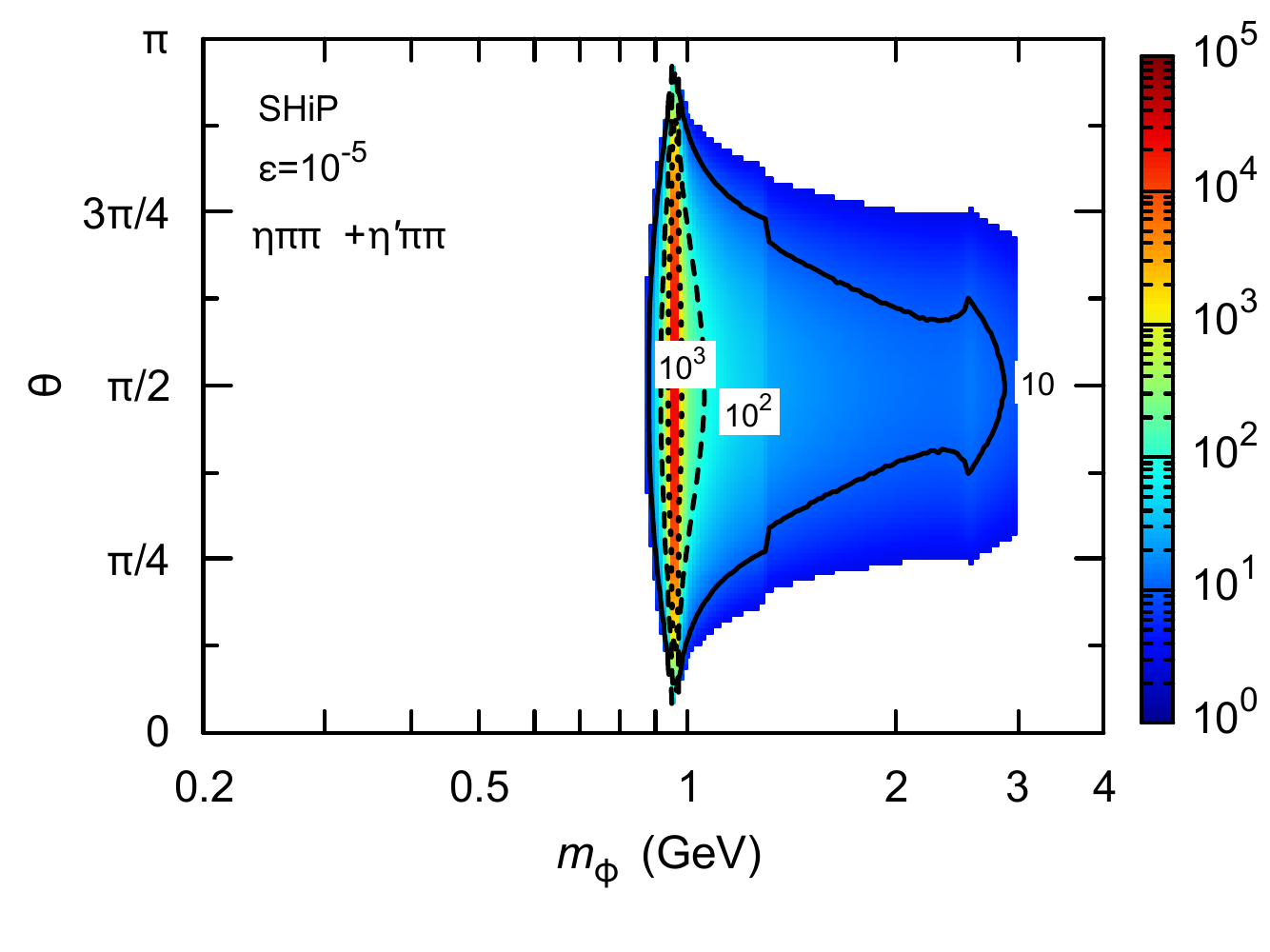} 
\end{tabular}
    \caption{Same plots as in Figure~\ref{fig:num-of-mesons-epsf1e-5-ship} for the individual final states indicated in each panel.}
\label{fig:num-of-mesons-epsf1e-5-ship-2}
\end{figure}
As shown in the left panel of Figure \ref{fig:num-of-mesons-epsf1e-5-ship}, for a fixed $m_\phi$, the expected number of two-meson decays first increases with $\theta$, reaches a maximum around $\theta=\pi/4$, and then vanishes at $\theta=\pi/2$. This behavior results from the interplay between the production rate and decay branching ratio of the dark scalar. As $\theta$ increases, more dark scalars are produced in $B$-meson decays, as shown in Figure~\ref{fig:meson-br}. At the same time, the branching ratios into two-meson final states decrease as $\theta$ approaches $\pi/2$. These decays vanish at $\theta=\pi/2$ since the dark scalar is purely CP-odd. 
The left panels of Figure~\ref{fig:num-of-mesons-epsf1e-5-ship-2} show the expected number of events of individual decay modes into two and four mesons.  
For $m_\phi < 1$ GeV, the $\pi\pi$ final states (left top) dominate the decays while for $m_\phi > 1$ GeV, $KK$ final states (left middle) dominate. The $\pi\pi\pi\pi,~\pi\pi KK$ and $\eta\eta$ final states (left bottom) are subdominant.
In the right panel of Figure \ref{fig:num-of-mesons-epsf1e-5-ship}, three-meson final states include $\pi\pi\pi$, $\pi KK$ and $\eta \pi\pi$, $\eta' \pi\pi$. These decays are forbidden at $\theta =0$. The expected number of the three-meson decays becomes maximum at around $\theta = \pi/2$. This is simply due to the increase of the production and decay branching ratios. The right panels of Figure~\ref{fig:num-of-mesons-epsf1e-5-ship-2} show the expected number of the individual decay modes into the three mesons.
The $\pi\pi\pi$ (right top) and $\eta\pi\pi$ (right bottom) final states show a sharp enhancement at $m_\phi = m_{\eta'}~(\simeq 0.96$ GeV) due to the mixing with the $\eta'$ meson. For $m_\phi > 1$ GeV, the $\pi KK$ final state (right middle) dominates and the $\eta \pi\pi + \eta'\pi\pi$ final state (right bottom) is subdominant.

\begin{figure}[t]
\includegraphics[scale=0.45]{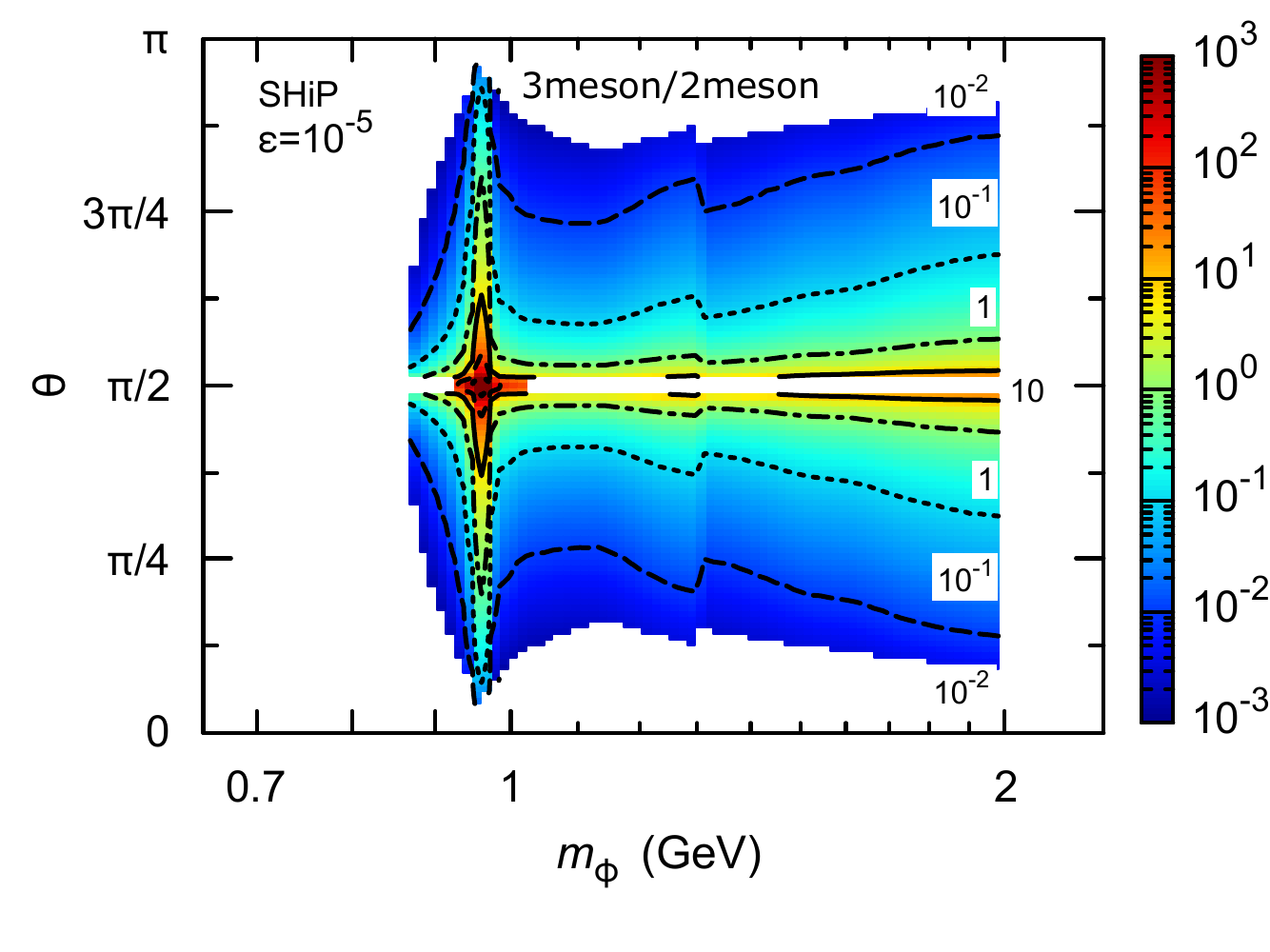} 
\caption{The ratio of three-meson events to two-meson events in the $(m_\phi, \theta)$ plane as a color map for $\epsilon= 10^{-5}$. The contours correspond to the ratio from $10^{-2}$ to $100$.}
\label{fig:num-of-meson-ratio-epsf1e-5-ship}
\end{figure}
Figure~\ref{fig:num-of-meson-ratio-epsf1e-5-ship} shows the ratio of three-meson events to two-meson events in the $(m_\phi, \theta)$ plane as a color map. 
In the figure, the contours are drawn for the ratios from $10^{-2}$ to $100$ ($m_\phi \simeq m_{\eta'}$ and $\theta\simeq \pi/2$). It was found that two and three-meson decays can be simultaneously observed over $\pi/8 \lesssim \theta \lesssim 7\pi/8$ for $m_\phi > 0.9$ GeV. The three-meson events can be larger than two-meson events for $\theta$ close to $\pi/2$ and/or $m_\phi = m_{\eta'}$. In other parameter regions, two-meson events are expected to be larger. 
These results suggest that the proposed strategy could probe CP violation under
the assumptions of our analysis.

\begin{figure}[t]
\hspace*{-1.0cm}
\centering
\begin{tabular}{cc}
    \includegraphics[scale=0.4]{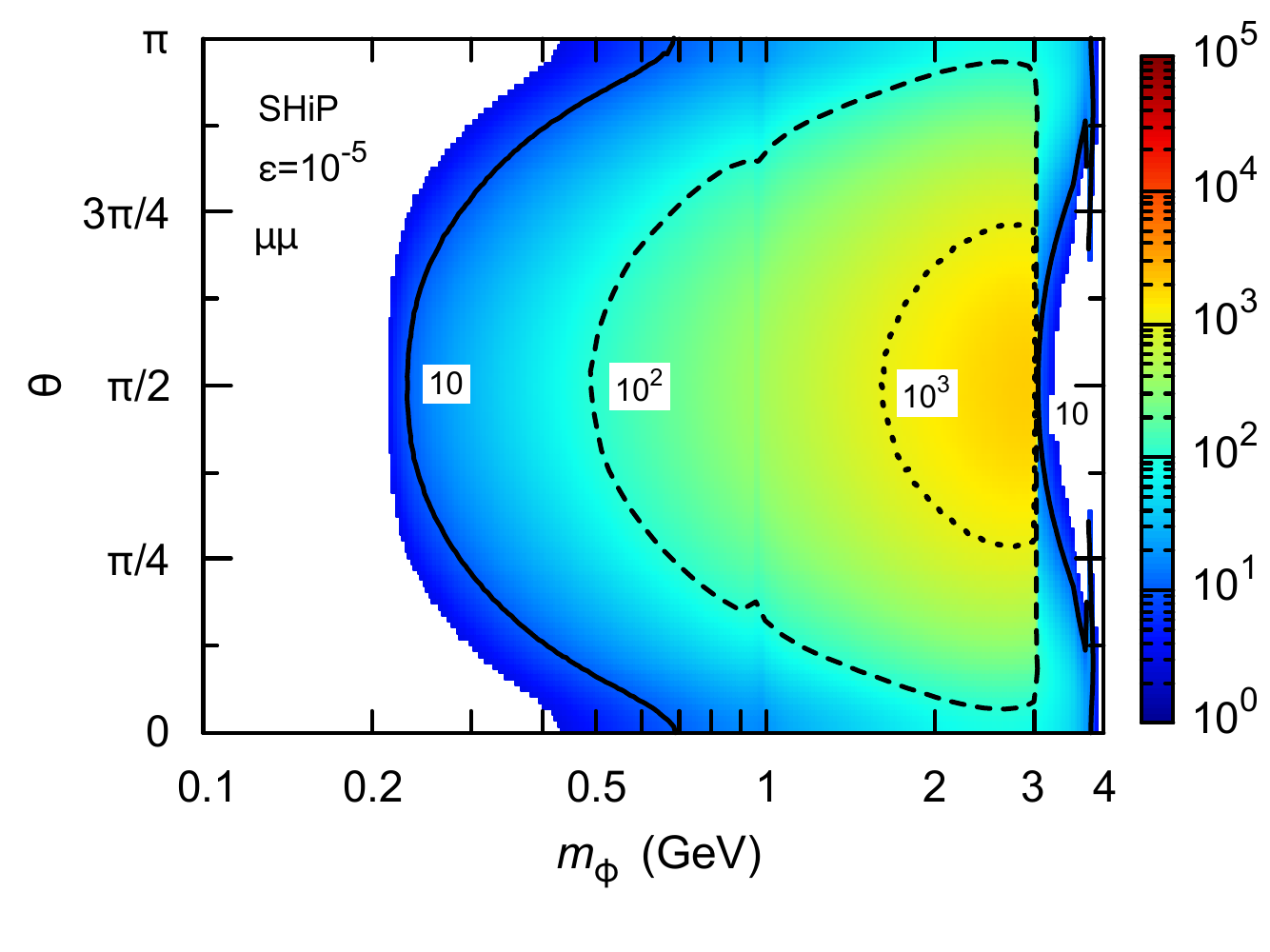} &
    \includegraphics[scale=0.4]{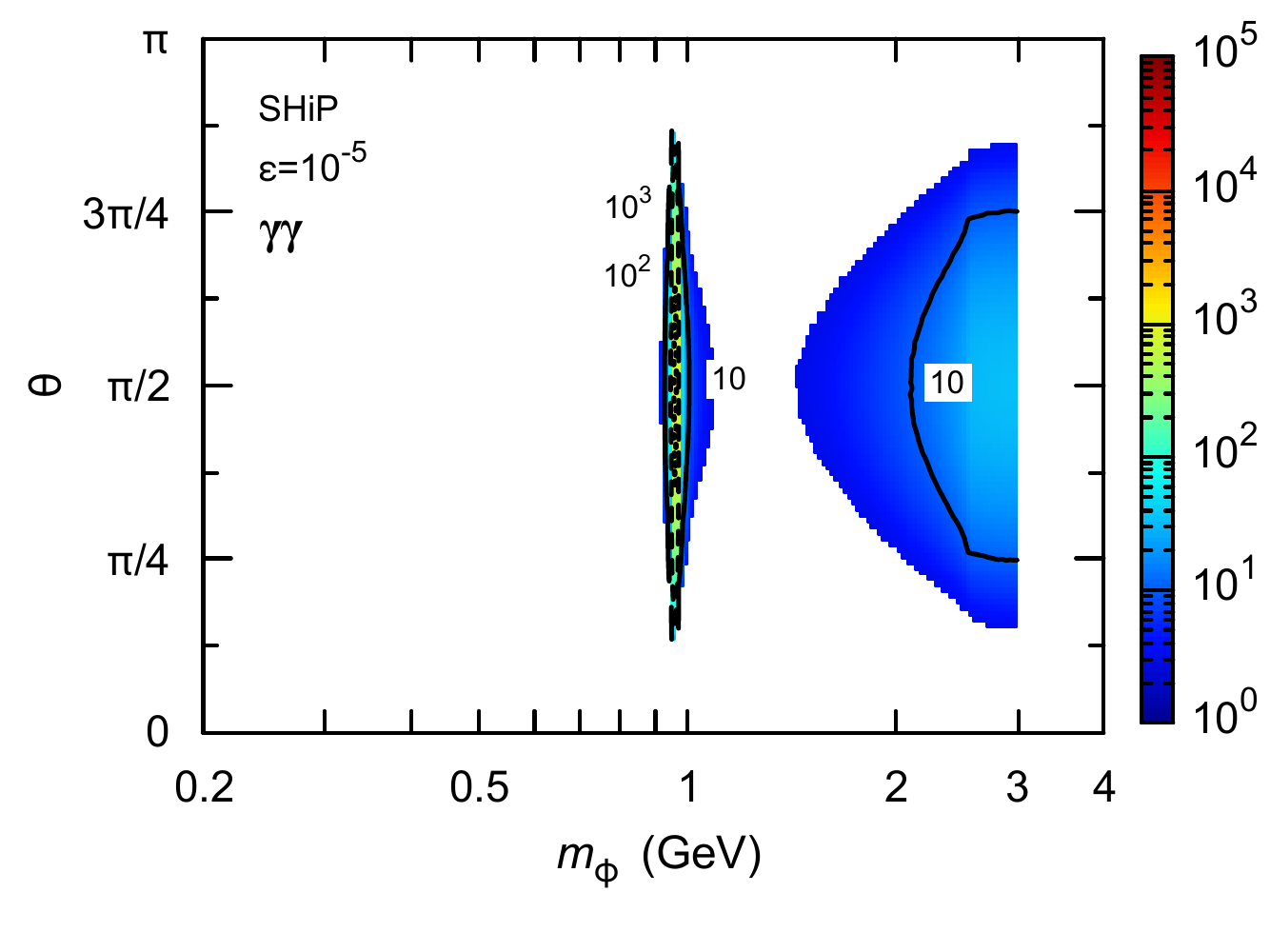} 
\end{tabular}
    \caption{Same plots as in Figure~\ref{fig:num-of-mesons-epsf1e-5-ship} for the decays into two-muon (Left) and di-photon (Right).
    }
\label{fig:num-of-2mu-gamgam-epsf1e-5}
\end{figure}

In contrast to the two-meson and three-meson final states, the two-muon and di-photon final states exist both in the pure CP-even and odd scalar decays. The two-muon decay width Eq.~\eqref{eq:decay-width-lepton}  is almost independent of $\theta$ for $m_\phi \gg m_\mu$, and can be a good reference for the CP violation search by comparing with the two-meson final states. On the other hand, although it is loop-induced and suppressed by $\alpha_{\rm em}^2$, the di-photon decay width Eq.~\eqref{eq:decay-width-2gamma} can depend on $\theta$ through the loop-induced couplings. This decay can be enhanced by the pseudoscalar meson mixing. Figure~\ref{fig:num-of-2mu-gamgam-epsf1e-5} shows the expected number of the dark scalar decays into two-muon (left) and di-photon (right) final states in the $(m_\phi, \theta)$ plane as a color map. 
%
%
\begin{figure}[t]
\hspace*{-1.0cm}
\centering
\begin{tabular}{cc}
    \includegraphics[scale=0.4]{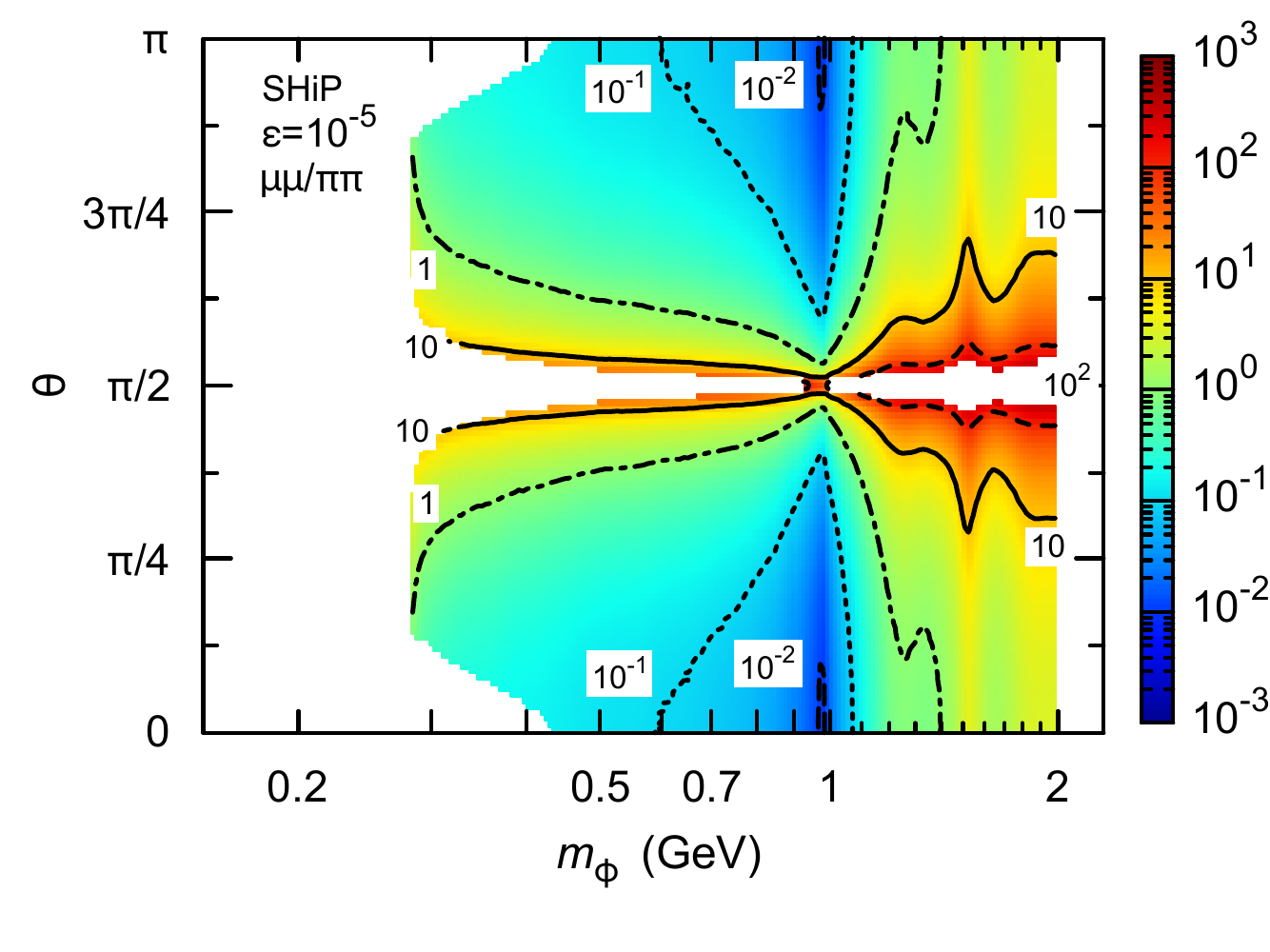} & 
    \includegraphics[scale=0.4]{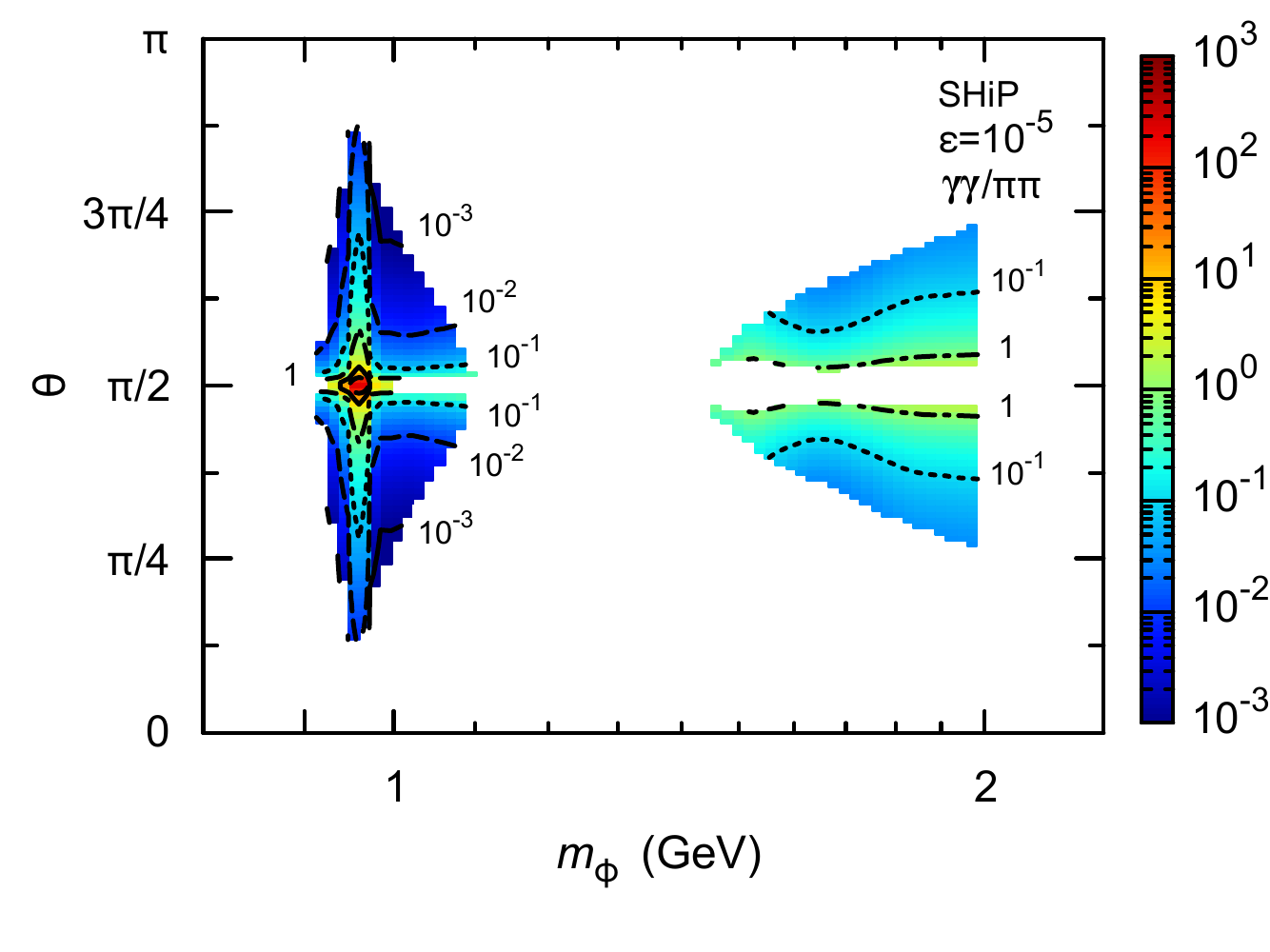}
\end{tabular}
    \caption{The ratio of two-muon (Left) and di-photon (Right) events to two-pion events in the $(m_\phi, \theta)$ plane as a color map for $\epsilon= 10^{-5}$.
    }
\label{fig:num-of-2mu-gamgam-ratio-epsf1e-5}
\end{figure}
%
The two-muon final state can be dominant as $\theta$ approaches $\pi/2$ due to the suppression of two-meson final states. 
The di-photon final states can be non-negligible for $m_\phi > 1$ GeV. 
The ratios of the two-muon (left) and di-photon (right) events to the $\pi\pi$ events are shown in Figure~\ref{fig:num-of-2mu-gamgam-ratio-epsf1e-5}. One can see in the left panel that the ratio of two-muon events to the $\pi\pi$ events shows good sensitivity to $\theta$ and covers a lighter mass range $m_\phi > 0.3$ GeV for almost the whole range of $\theta$. Thus, when the dark scalar boson is lighter than $0.9$ GeV, this ratio could be used for the CP violation search. The ratio of the di-photon events to the $\pi\pi$ events also shows the $\theta$ dependence around $m_\phi = m_{\eta'}$ and $> 1.5$ GeV.
For $m_\phi > 0.9$ GeV, these decays can provide complementary information for the CP violation search combined with the two-meson and three-meson final states.

\begin{figure}[t]
\hspace*{-1.0cm}
\centering
\begin{tabular}{cc}
    \includegraphics[scale=0.4]{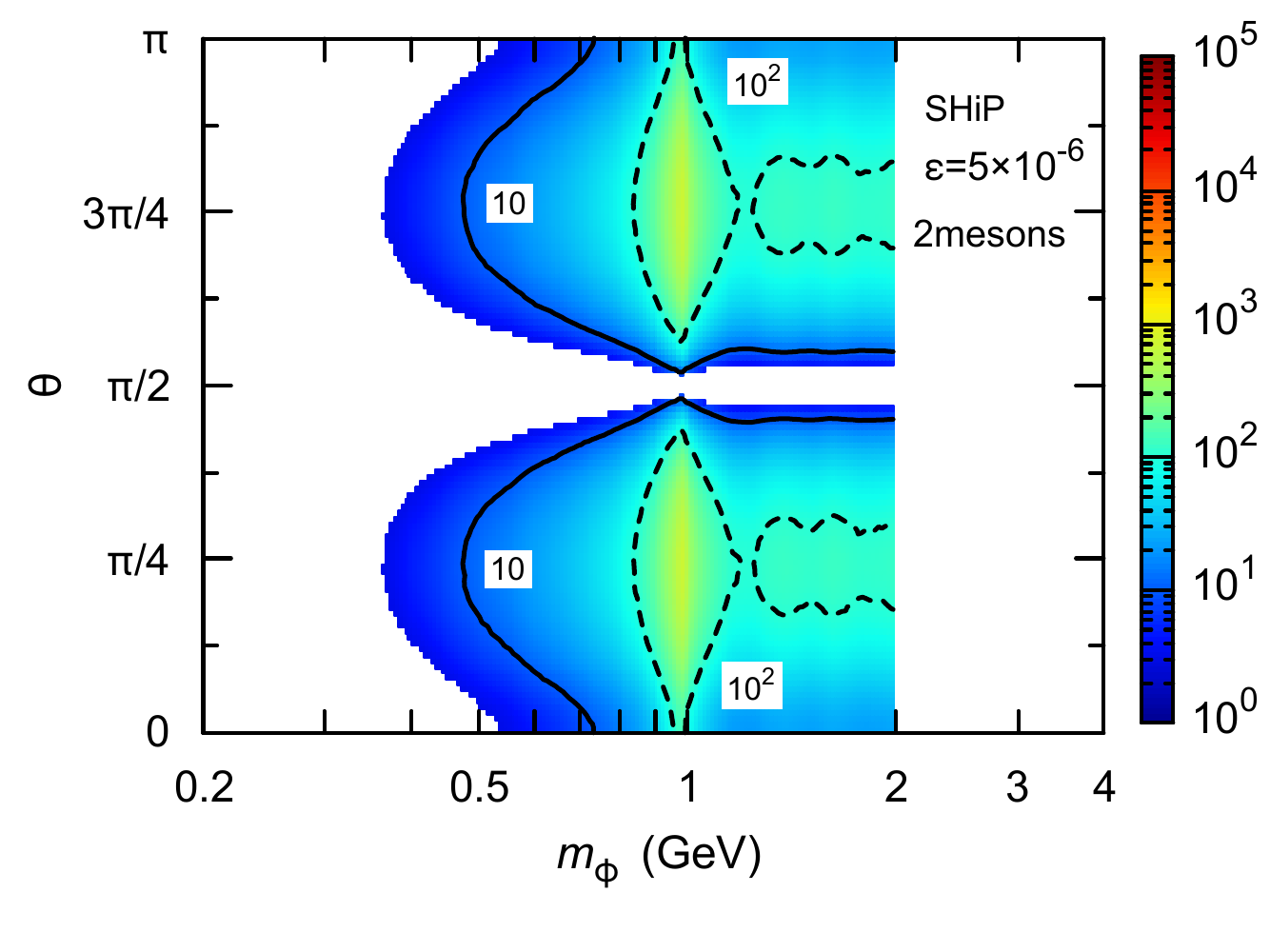} &
    \includegraphics[scale=0.4]{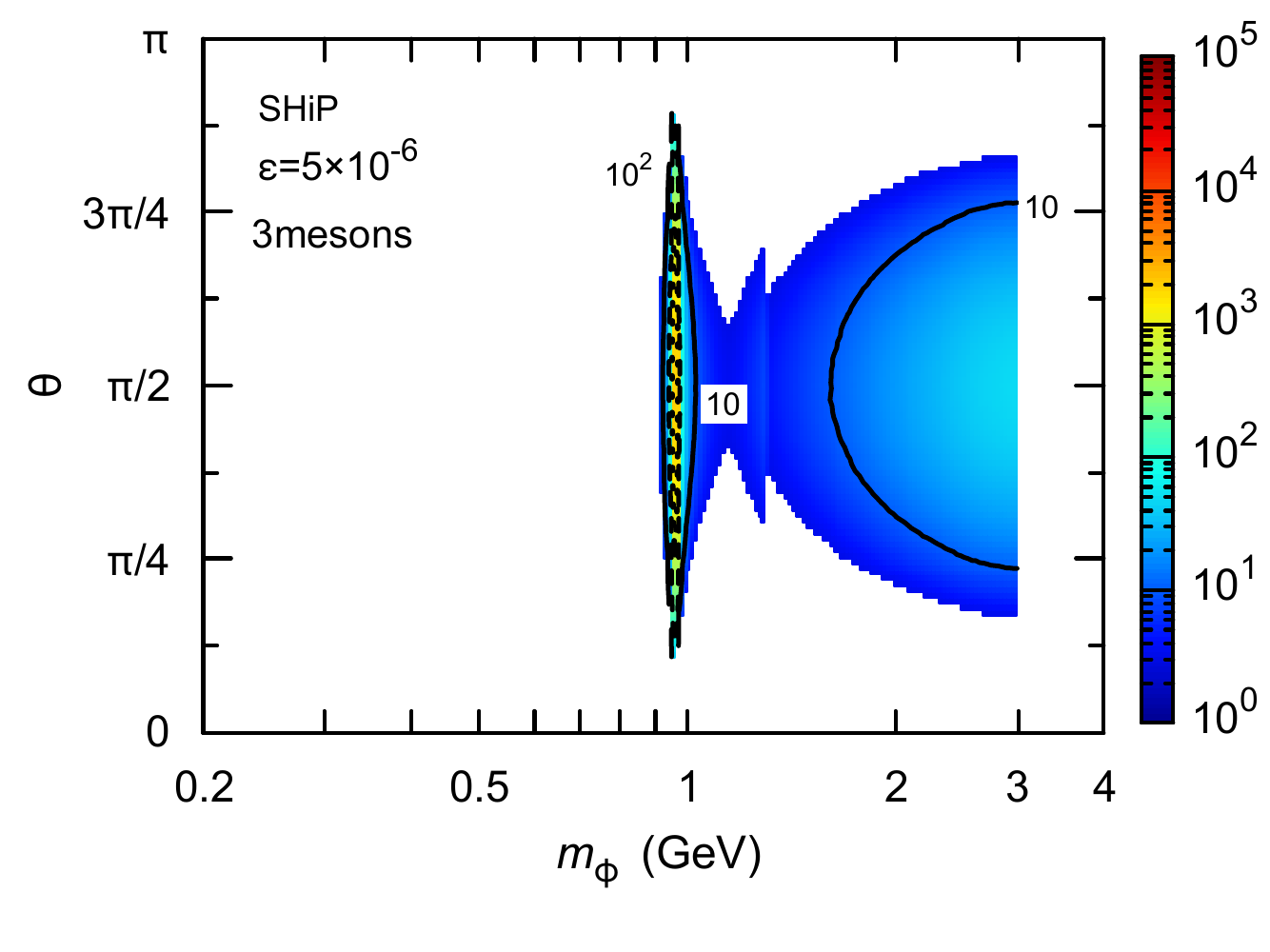} 
\end{tabular}
    \caption{The same plots as in Figure~\ref{fig:num-of-mesons-epsf1e-5-ship} for $\epsilon = 5 \times 10^{-6}$.
    }
\label{fig:num-of-mesons-epsf5e-6-ship}
\end{figure}
\begin{figure}[t]
\includegraphics[scale=0.45]{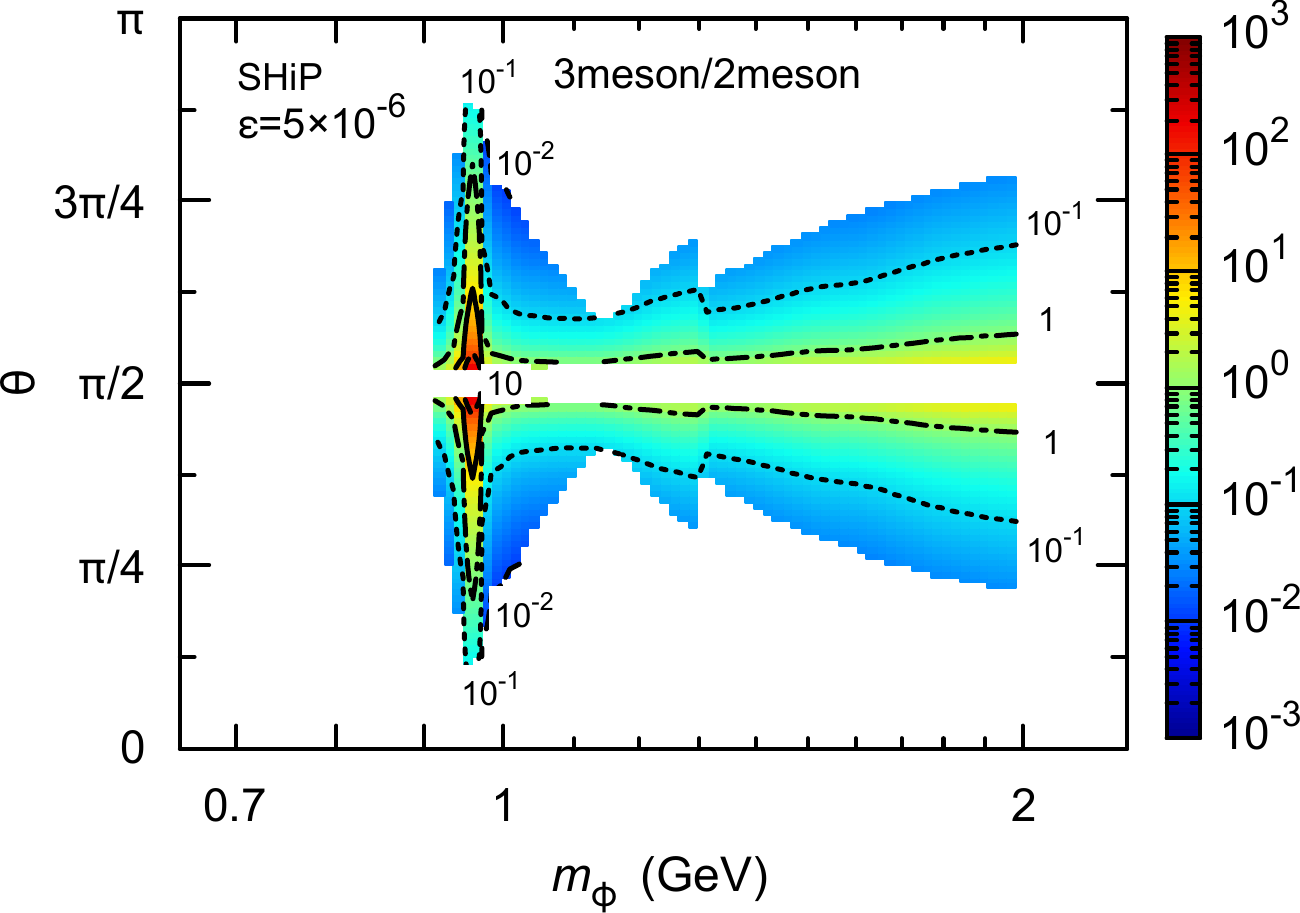} 
\caption{The same plots as in Figure~\ref{fig:num-of-meson-ratio-epsf1e-5-ship} for $\epsilon = 5 \times 10^{-6}$.
}
\label{fig:num-of-meson-ratio-epsf5e-6-ship}
\end{figure}
For comparison, in Figure~\ref{fig:num-of-mesons-epsf5e-6-ship}, 
we show the same plots as in Figure~\ref{fig:num-of-mesons-epsf1e-5-ship} for $\epsilon = 5 \times 10^{-6}$. Although the expected number of events is reduced by a factor of $4$ compared with that for $\epsilon = 10^{-5}$, the predicted numbers remain above the three events.
Figure~\ref{fig:num-of-meson-ratio-epsf5e-6-ship} shows the ratio of three-meson events to two-meson events in the $(m_\phi, \theta)$ plane as a color map for $\epsilon= 5 \times 10^{-6}$. The ratio is between $10^{-2}$ and $10$ for $m_\phi > 0.9$ GeV and $\pi/4 \lesssim \theta \lesssim 3\pi/4$. Similar to Figure~\ref{fig:num-of-meson-ratio-epsf1e-5-ship}, 
CP violation could be probed in this region under the same assumptions.

\begin{figure}[t]
\hspace*{-1.0cm}
\centering
\begin{tabular}{cc}
    \includegraphics[scale=0.4]{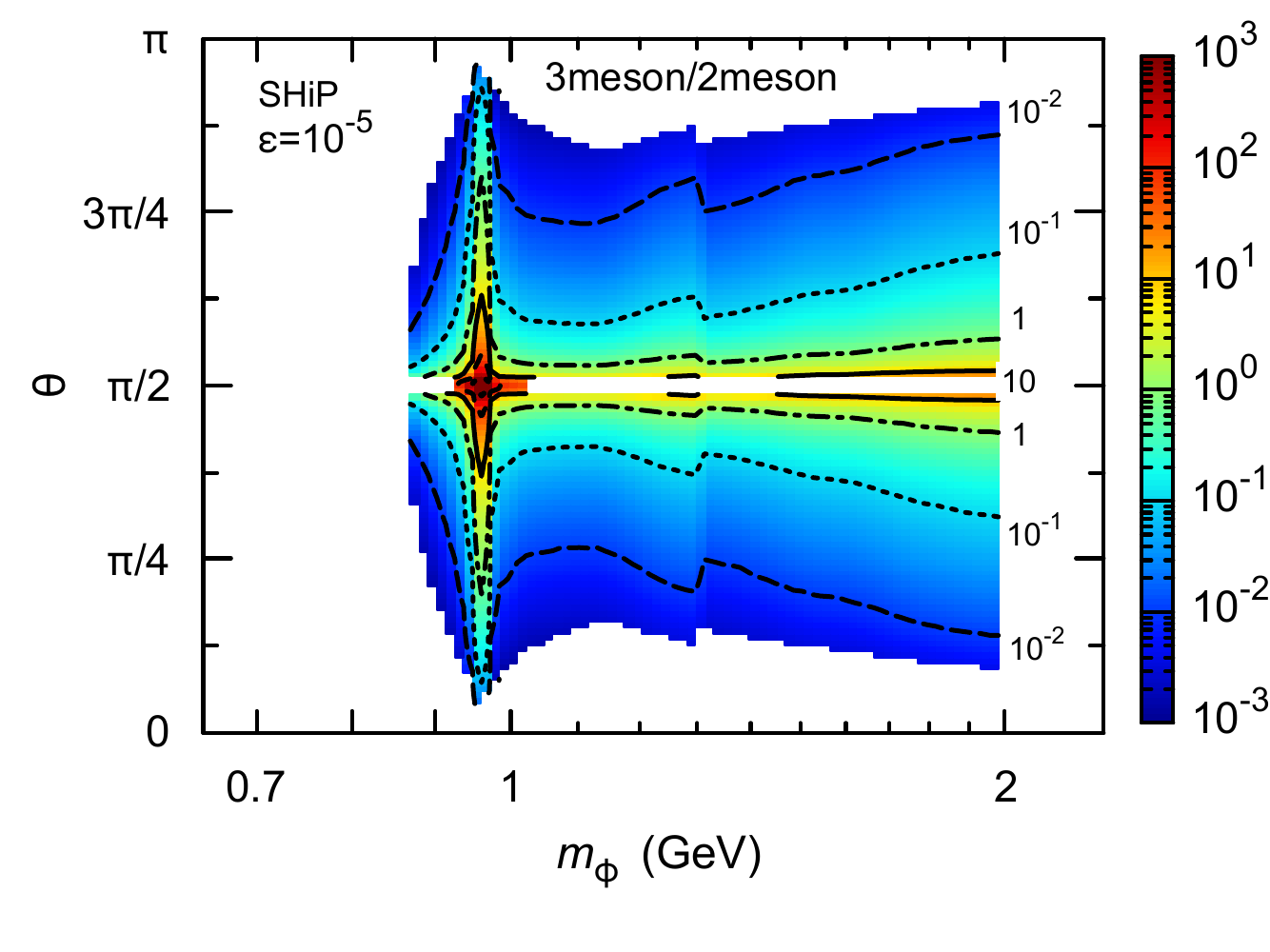} &
    \includegraphics[scale=0.4]{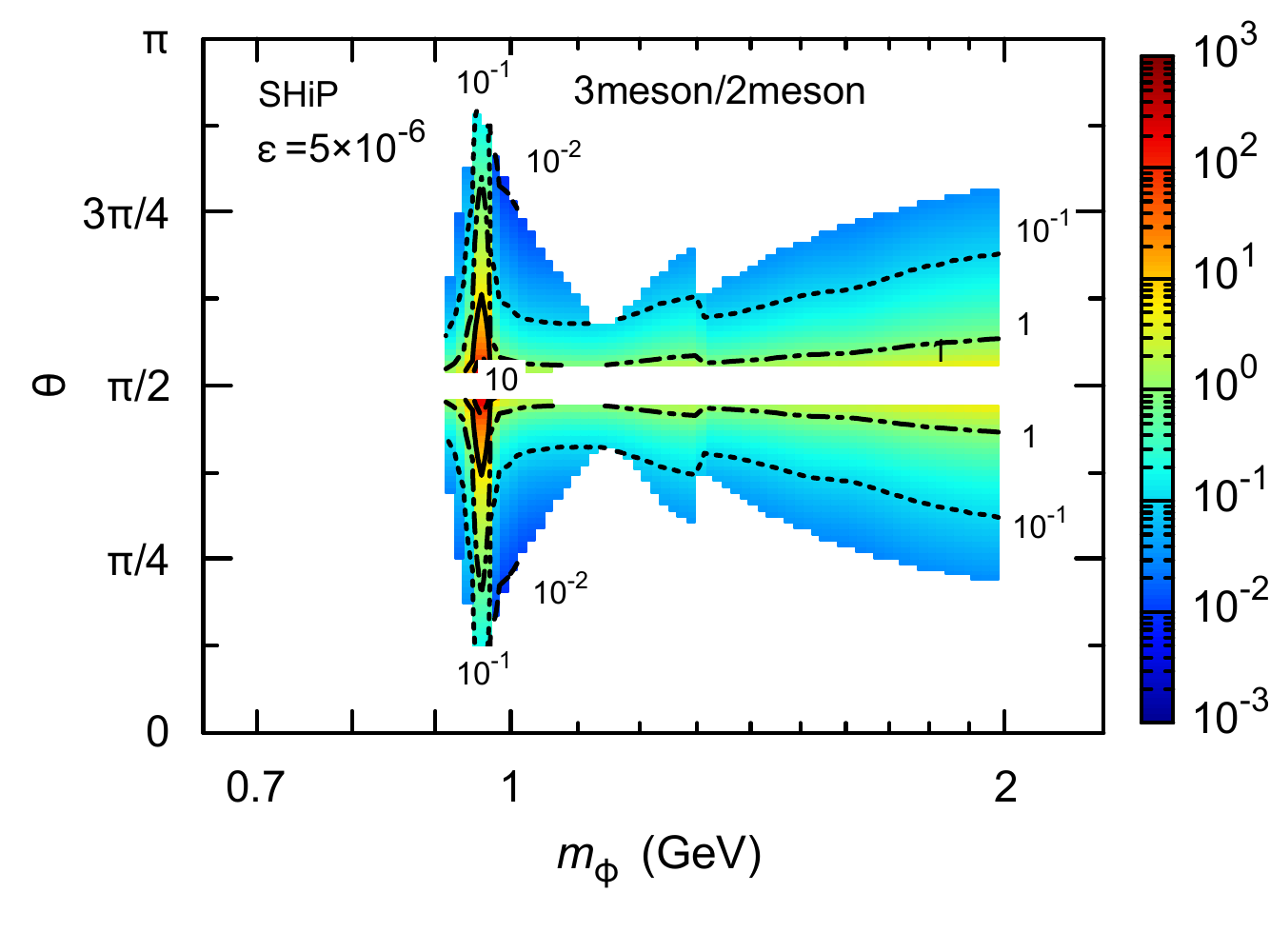} 
\end{tabular}
    \caption{The same plots as in Figure~\ref{fig:num-of-meson-ratio-epsf1e-5-ship} for the case where $\epsilon$ for the dark scalar couplings to leptons is set to zero. For the couplings to quarks, $\epsilon = 10^{-5}$ (Left) and $5 \times 10^{-6}$ (Right). }
\label{fig:num-of-meson-ratio-epsq1e-5-ship}
\end{figure}
As shown in Figure~\ref{fig:branching-ratio}, the branching ratio into the two-muon final state is sizable. Thus, for a leptophobic dark scalar, the total decay width is reduced and the decay length becomes longer, which may affect the expected number of signal events.
To examine this effect, we set the couplings to leptons to zero and show in Figure~\ref{fig:num-of-meson-ratio-epsq1e-5-ship} the same plots as in Figure~\ref{fig:num-of-meson-ratio-epsf1e-5-ship} for $\epsilon = 10^{-5}$ (left) and $5 \times 10^{-6}$ (right).
Comparing Figures~\ref{fig:num-of-meson-ratio-epsf1e-5-ship} and \ref{fig:num-of-meson-ratio-epsf5e-6-ship} with Figure~\ref{fig:num-of-meson-ratio-epsq1e-5-ship}, we find that the ratio of the three-meson to two-meson events remains almost unchanged, despite the modification of the total decay width.

\subsubsection{FASER2}
\begin{figure}[t]
\hspace*{-1.0cm}
\centering
\begin{tabular}{cc}
    \includegraphics[scale=0.4]{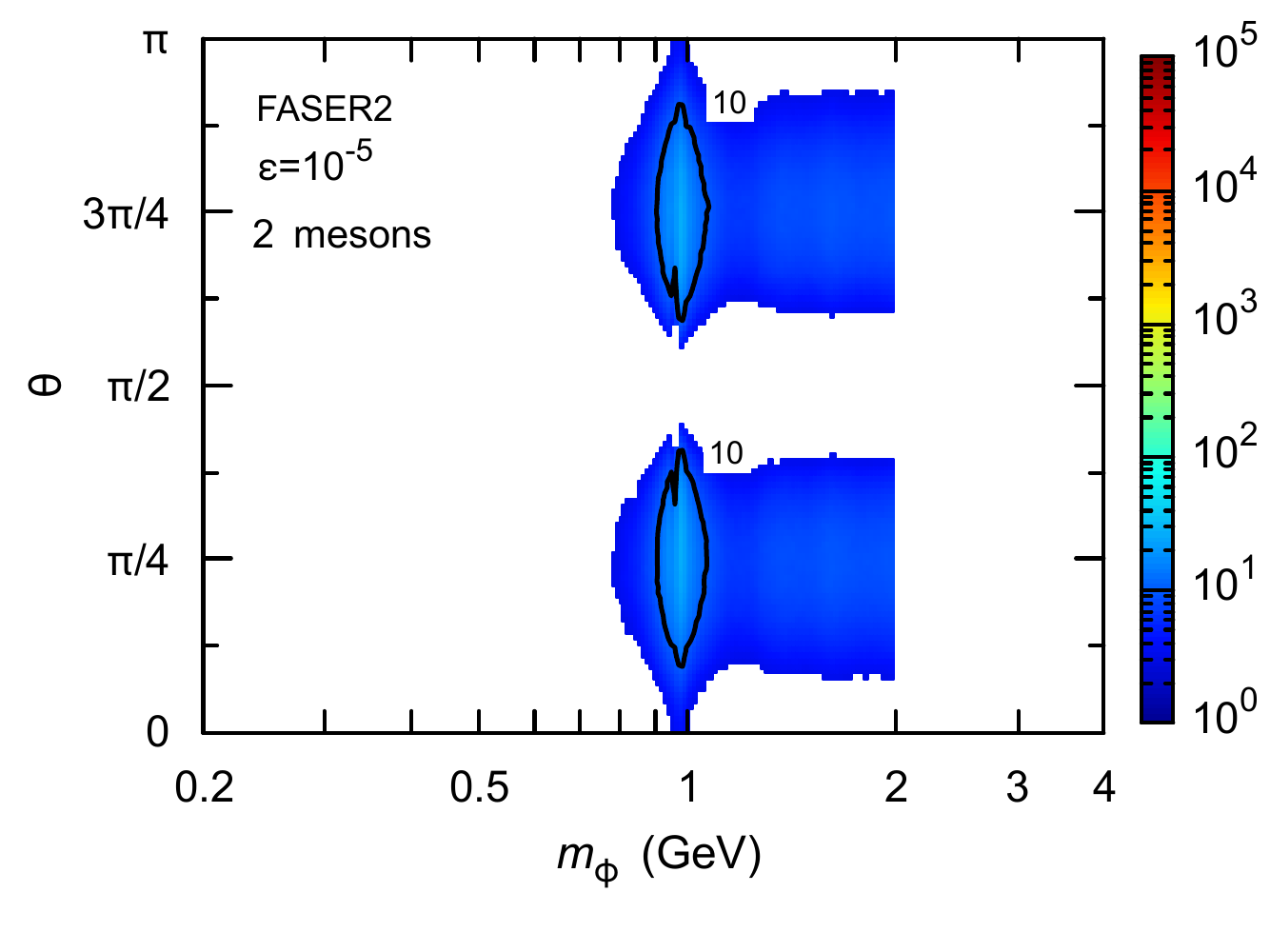} &
    \includegraphics[scale=0.4]{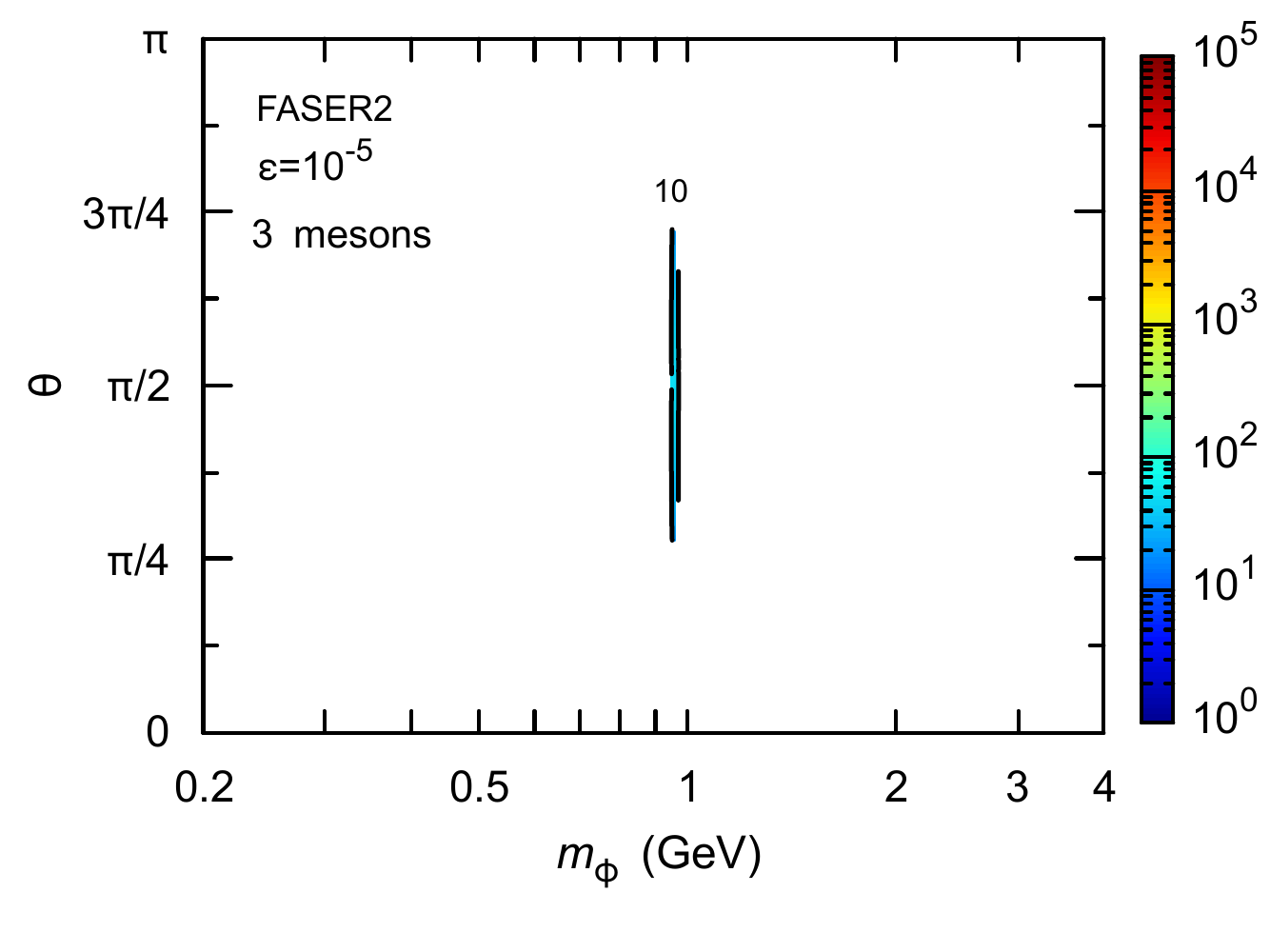} 
\end{tabular}
    \caption{Number of dark scalar decays into two-meson (Left) and three-meson (Right) final states at FASER2.}
\label{fig:num-of-mesons-epsf1e-5-faser2}
\end{figure}
%
\begin{figure}[t]
\includegraphics[scale=0.4]{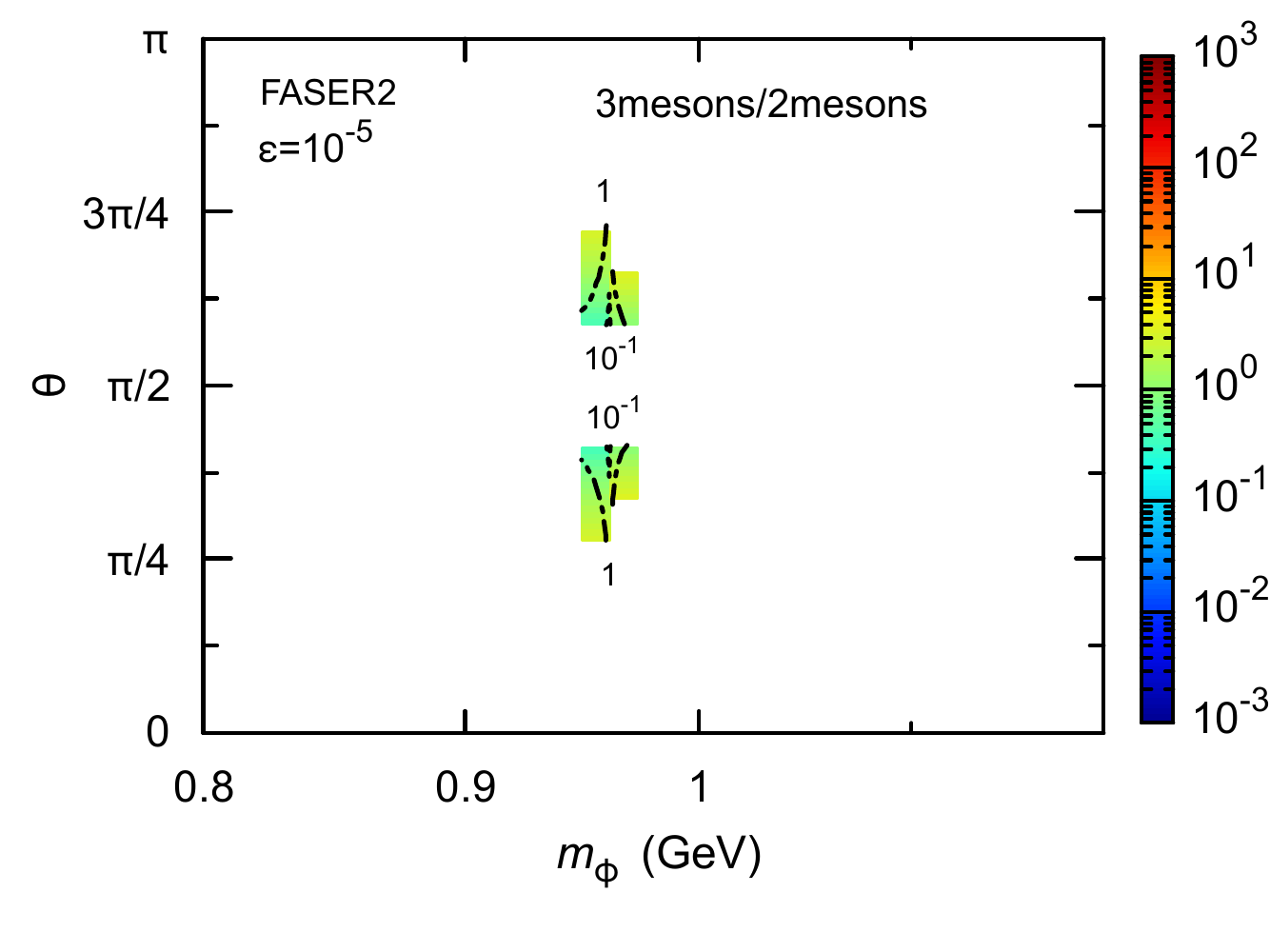} 
\caption{The ratio of three-meson events to two-meson events in the $(m_\phi, \theta)$ plane as a color map for $\epsilon= 10^{-5}$ at FASER2.}
\label{fig:num-of-meson-ratio-epsf1e-5-faser2}
\end{figure}
%
At the end of this section, we show the results for the FASER2 experiment for comparison. Figures~\ref{fig:num-of-mesons-epsf1e-5-faser2} and \ref{fig:num-of-meson-ratio-epsf1e-5-faser2} are the same plots as Figures~\ref{fig:num-of-mesons-epsf1e-5-ship} and \ref{fig:num-of-meson-ratio-epsf1e-5-ship} for the FASER2 experiment. 
The detector geometry and the integrated luminosity are given in
Table~\ref{tab:detector-dimension}. Although the expected number of events is
smaller than that at the SHiP experiment due to the smaller detector size of
FASER2, it remains above the three events for two-meson decays in $m_{\eta'} \leq m_\phi \leq 2~\mathrm{GeV}$.
Three-meson final states are also predicted 
to produce up to about $10$ events for $m_\phi \simeq 1$ GeV and $\pi/4 \lesssim \theta \lesssim 3\pi/4$.
Requiring both two-meson and three-meson events to be larger than three, the ratio is found to range from $10^{-1}$ to $1$ at
$m_\phi \approx m_{\eta'}$ for $\pi/4 \lesssim \theta \lesssim 3\pi/8$ and $5\pi/8 \lesssim \theta \lesssim 3\pi/4$.

\section{Summary} \label{sec:summary}
We have considered the effective interactions of the CP-violating dark scalar with the SM particles. We proposed a strategy to probe the CP nature of the dark scalar by comparing the two-meson and three-meson final states in the dark scalar decays. We have analyzed the expected number of events at the SHiP and FASER2 experiments. 

For the SHiP experiment, it was found that a sufficient number of dark scalar decays can be expected for $\epsilon = 10^{-5}$ and $5 \times 10^{-6}$ in the mass range of $m_\phi \geq 0.9$ GeV. 
The $KK$ final states are dominant in the two-meson decays while the $\pi KK$ final states are dominant in the three-meson ones. 
It was also shown that the ratio of three-meson events to two-meson events 
is sensitive to the CP-violating phase, varying from $10^{-2}$ to $100$ over the range of
$\pi/8 \lesssim \theta \lesssim 7\pi/8$ for $\epsilon = 10^{-5}$.
 The ratios of two-muon and di-photon events to $\pi\pi$ events also show the CP-violating phase dependence and can provide complementary probes of CP violation when combined with the two-meson and three-meson final states. We have obtained similar results for the case of $\epsilon = 5 \times 10^{-6}$ and for the model in which only quarks couple to the dark scalar. For the FASER2 experiment, it was found that the expected number of events is smaller than that of the SHiP experiment due to the smaller detector size. The ratio of three-meson events to two-meson events is found to range from $10^{-1}$ to $1$ at $m_\phi \approx m_{\eta'}$ for $\pi/4 \,(5\pi/8)\lesssim \theta \lesssim 3\pi/8 \,(3\pi/4)$.

Once the dark scalar boson is discovered, determining its CP nature will be one of the primary questions in understanding the physics of the dark sector. Our analysis is based on the effective interactions of the dark scalar boson with the SM particles. Such interactions can be generated by some classes of models similar to the one presented in the Appendix~\ref{apdx:UV-complete-model}. However, the same form of the interactions can be obtained in extended Higgs models such as two Higgs doublet models with a CP-violating Higgs potential. Thus, our analysis is applicable to a wide class of models.  
It is also worth noting that our study is useful for distinguishing a CP-violating dark scalar from a dark photon, or more generally, from a dark vector boson. The dark vector boson can have the same decay final states when it couples to the light quarks. Such dark vector decays can mimic the CP-violating dark scalar decays. We will discuss the possibility of distinguishing these dark particles at future LLP experiments in our future work.

\section*{Acknowledgements}
This work was partially supported by JSPS KAKENHI Grant Numbers JP23K03384 (M.A.) 
and JP22K03622, JP26K07108, JP26H02052 (T.S.).

\appendix

\section{A UV-complete model} \label{apdx:UV-complete-model}
We present a simple renormalizable UV-complete model that realizes the effective interaction Lagrangian Eq.~\eqref{eq:lag-int}. We first focus on the scalar and pseudoscalar interactions of the dark scalar boson with the SM fermions. We then show how the interactions with the electroweak gauge bosons can arise through mixing between the dark scalar boson and the SM Higgs boson. In this appendix, we assume the minimal flavor violation hypothesis \cite{DAmbrosio:2002vsn} to avoid flavor changing neutral current constraints.

To begin with, we introduce a complex SM-singlet scalar,
\begin{align}
S=\frac{1}{\sqrt{2}}(\phi+ia),
\end{align}
together with heavy vector-like fermions. For illustration purposes, let us focus on a down-type quark $d$ and introduce a vector-like quark with the same quantum charges as the right-handed $d$ quark,
\begin{align}
D_{L,R} \sim ({\bf 3},{\bf 1},-1/3).
\end{align}
The relevant renormalizable interaction Lagrangian is given by
\begin{align}
\mathcal{L} \supset & -y_d \overline{Q}_L H d_R 
- M_D \overline{D}_L D_R 
- \lambda_Q \overline{Q}_L H D_R -\lambda_d \overline{D}_L d_R S+\mathrm{h.c.}
\end{align}
where $M_D$ is the mass of the vector-like quark and $\lambda_Q$ and $\lambda_d$ are the Yukawa couplings. For $M_D$ much heavier than the electroweak scale, integrating out the vector-like quark generates the effective interaction term, 
\begin{align}
\mathcal{L}_{\rm eff} \supset \frac{c_d}{\Lambda} \overline{Q}_L H d_R S + \mathrm{h.c.},
\end{align}
where
\begin{align}
\frac{c_d}{\Lambda} = \frac{\lambda_Q\lambda_d}{M_D}.
\end{align}
Similar interactions can be generated for up-type quarks and charged leptons by introducing the corresponding vector-like fermions.

After electroweak symmetry breaking, the effective interaction term involving the real part of $S$ can be written as
\begin{align}
\mathcal{L}_{\rm eff} \supset -\frac{m_f}{v} 
\overline{f} ( \xi^s_f + i \xi^p_f \gamma_5 ) f \phi,
\end{align}
where the Yukawa couplings are factored out as a normalization. The modifiers $\xi^s_f$ and $\xi^p_f$ are determined by the magnitude and the physical phase of $c_f$. The complex phase of $c_f$ can be physical when the scalar potential contains another source of CP violation that fixes the phase convention of $S$.

\section{Loop functions} \label{apdx:loop-function}
The modifier parameters of the dark scalar interactions with the photon and gluon at one-loop order are given by \cite{Fugel:2004ug,Djouadi:2005gi,Domingo:2016yih}
\begin{subequations}
    \begin{align}
        \xi_{\gamma} &= N_c \sum_f Q_f^2 \xi^s_{f} \mathcal{A}_{1/2} (\tau_f), \\
        \xi_{g} &= \frac{3}{2} \sum_f \xi^s_{f} \mathcal{A}_{1/2} (\tau_f), \\
        \tilde{\xi}_{\gamma} &= \frac{1}{2} N_c \sum_f Q_f^2 \xi^p_{f} \tilde{\mathcal{A}}_{1/2} (\tau_f), \\
        \tilde{\xi}_{g} &= \frac{1}{4} \sum_f \xi^p_{f} \tilde{\mathcal{A}}_{1/2} (\tau_f),
    \end{align} \label{eq:loop-induced-coup}
\end{subequations}
where $\tau_i = m_{\phi}^2/4 m_i^2$ with $i=f$ running over the fermions in the loops. The number of color degrees of freedom $N_c$ is $1$ for leptons or $3$ for quarks, respectively, and the electric charge of the fermion $Q_f$ is normalized by $e = \sqrt{4\pi \alpha_{\rm em}}$. 

\begin{subequations}
    \begin{align}
        \mathcal{A}_{1/2}(\tau) &= [\tau + (\tau-1) f(\tau)]\tau^{-2}, \\
        \tilde{\mathcal{A}}_{1/2} (\tau) &= 2 \tau^{-1} f(\tau),
    \end{align}
\end{subequations}
with
\begin{align}
    f(\tau) = 
    \begin{cases}
        (\arcsin\sqrt{\tau})^{2} & \mathrm{for}~\tau \leq 1 \\
        -\frac{1}{4} \left( \log \frac{1 + \sqrt{1-1/\tau}}{1 - \sqrt{1-1/\tau}} - i \pi \right)^2 & \mathrm{for}~\tau > 1
    \end{cases}.
\end{align}

\section{Lagrangian of chiral anomaly} \label{apdx:chiral-anomaly}
\begin{align}
    \mathcal{L}_{\mathrm{axial~anomaly}} = 
    \frac{C}{2} (\partial_\mu K^\mu)^2 + i (\partial_\mu K^\mu) \left[ \frac{1}{4} \mathrm{Tr}\log \Sigma - \frac{1}{4} \mathrm{Tr}\log \Sigma^\dagger - i \tilde{\xi}_{g} \frac{\hat{\phi}}{v} \right], \label{eq:lag-axial-anom}
\end{align}
where $C$ is a constant parameter and $\partial_\mu K^\mu \sim \frac{\alpha_s}{4\pi} G^a_{\mu\nu} \tilde{G}^{a\mu\nu}$ is an auxiliary field. Eliminating the auxiliary field \(\partial_\mu K^\mu\) through its equation of motion, this Lagrangian becomes
\begin{align}
    \mathcal{L}_{\mathrm{axial~anomaly}} = \frac{1}{2C} \left[ \frac{1}{4} \mathrm{Tr}\log \Sigma - \frac{1}{4} \mathrm{Tr}\log \Sigma^\dagger - i \tilde{\xi}_{g} \frac{\hat{\phi}}{v} \right]^2. 
\end{align}

\section{Interaction Lagrangian in mass eigenstates} \label{apdx:meson-int-masseigenstates}
The mass terms of the dark scalar boson and mesons are read from Eqs.~\eqref{eq:lagrangian-chipt} and \eqref{eq:lag-axial-anom} as
\begin{align}
\mathcal{L}_{\rm mass} &= \frac{B}{2}\,\mathrm{Tr} \left[M_q[\hat{\phi}] \Sigma+\Sigma^\dagger M_q^\dagger[\hat{\phi}] \right]
+ \frac{1}{2C} \left[ \frac{1}{4} \mathrm{Tr}\log \Sigma - \frac{1}{4} \mathrm{Tr}\log \Sigma^\dagger - i \tilde{\xi}_{g} \frac{\hat{\phi}}{v} \right]^2,
\end{align}
where the auxiliary field is integrated out using the minimization condition, and $\Sigma$ and $M_q[\hat{\phi}]$ are given by Eqs.~\eqref{eq:sigma} and \eqref{eq:field-dep-mass}, respectively. We denote $\hat{\phi}$ to distinguish its mass eigenstate in this appendix.
The $3\times 3$ meson matrices $\lambda$ are defined as 
\begin{subequations}
\begin{align}
\lambda^1 &= \frac{1}{\sqrt{2}}
\begin{pmatrix}
0 & 1 & 0 \\
1 & 0 & 0 \\
0 & 0 & 0
\end{pmatrix}, \quad
\lambda^2 = \frac{1}{\sqrt{2}}
\begin{pmatrix}
0 & -i & 0 \\
i & 0 & 0 \\
0 & 0 & 0
\end{pmatrix}, \quad
\lambda^3 = \frac{1}{\sqrt{2}}
\begin{pmatrix}
1 & 0 & 0 \\
0 & -1 & 0 \\
0 & 0 & 0
\end{pmatrix}, \\
\lambda^4 &= \frac{1}{\sqrt{2}}
\begin{pmatrix}
0 & 0 & 1 \\
0 & 0 & 0 \\
1 & 0 & 0
\end{pmatrix}, \quad
\lambda^5 = \frac{1}{\sqrt{2}}
\begin{pmatrix}
0 & 0 & -i \\
0 & 0 & 0 \\
i & 0 & 0
\end{pmatrix}, \quad
\lambda^6 = \frac{1}{\sqrt{2}}
\begin{pmatrix}0 & 0 & 0 \\
0 & 0 & 1 \\
0 & 1 & 0
\end{pmatrix}, \\
\lambda^7 &= \frac{1}{\sqrt{2}}
\begin{pmatrix}0 & 0 & 0 \\
0 & 0 & -i \\
0 & i & 0
\end{pmatrix}, \quad
\lambda^8 = \frac{1}{\sqrt{6}}
\begin{pmatrix}1 & 0 & 0 \\
0 & 1 & 0 \\
0 & 0 & -2
\end{pmatrix}, \quad
\lambda^9 = \frac{1}{\sqrt{3}}
\begin{pmatrix}
1 & 0 & 0 \\
0 & 1 & 0 \\
0 & 0 & 1
\end{pmatrix}.
\end{align}
\end{subequations}

The mass eigenstates of the pions, kaons and etas in the absence of the dark scalar boson, and the corresponding matrices are given by \cite{Domingo:2016yih}
\begin{align}
\hat{\pi}^0 &= \pi_3, ~~~~~~~~~~~~~~~~~~~ \lambda^{\hat{\pi}^0} = \lambda^3, \\
\hat{\pi}^\pm &= \frac{1}{\sqrt{2}}(\pi_1 \mp i \pi_2), ~~~~ \lambda^{\hat{\pi}^\pm} = \frac{1}{\sqrt{2}}(\lambda^1 \pm i \lambda^2), \\
\hat{K}^\pm &= \frac{1}{\sqrt{2}}(\pi_4 \mp i \pi_5), ~~~~ \lambda^{\hat{K}^\pm} = \frac{1}{\sqrt{2}}(\lambda^4 \pm i \lambda^5), \\
\hat{K}^0 &= \frac{1}{\sqrt{2}}(\pi_6 - i \pi_7), ~~~~ \lambda^{\hat{K}^0} = \frac{1}{\sqrt{2}}(\lambda^6 + i \lambda^7), \\
\bar{\hat{K}}^0 &= \frac{1}{\sqrt{2}}(\pi_6 + i \pi_7),~~~~~ \lambda^{\bar{\hat{K}}^0} = \frac{1}{\sqrt{2}}(\lambda^6 - i \lambda^7),\\
\hat{\eta} &= \cos\theta_\eta \pi_8 - \sin\theta_\eta \pi_9, ~~~~ \lambda^{\hat{\eta}} = \cos\theta_\eta \lambda^8 - \sin\theta_\eta \lambda^9,\\
\hat{\eta}' &= \sin\theta_\eta \pi_8 + \cos\theta_\eta \pi_9, ~~~~ \lambda^{\hat{\eta}'} = \sin\theta_\eta \lambda^8 + \cos\theta_\eta \lambda^9,
\end{align}
where $\theta_\eta$ is the $\eta$--$\eta'$ mixing angle and $\mathrm{Tr}(\lambda^i (\lambda^j)^\dagger) = \delta^{ij},~(i,j=\hat{\pi}^0,\hat{\pi}^\pm,\hat{K}^\pm,\hat{K}^0,\bar{\hat{K}}^0,\hat{\eta},\hat{\eta}')$. 
Then, adding the dark scalar mass term $-\frac{1}{2} m_{\phi}^2 \hat{\phi}^2$, the mass matrix can be written in the basis of  $(\hat{\pi}^0, \hat{\eta}, \hat{\eta}', \hat{\phi})$ as
\begin{align}
M^2_{\rm mass} &= 
O^T_\eta
\begin{pmatrix}
m_{\hat{\pi}^0}^2 & \frac{1}{\sqrt{3}} \delta & \sqrt{\frac{2}{3}} \delta & \delta m_3^2 \\
\frac{1}{\sqrt{3}} \delta & m_{\pi_8}^2 & \Delta & \delta m_8^2 \\
\sqrt{\frac{2}{3}} \delta & \Delta & m_{\pi_9}^2 & \delta m_9^2 \\
\delta m_3^2 & \delta m_8^2 & \delta m_9^2 & \bar{m}_{\hat{\phi}}^2
\end{pmatrix} 
O_\eta,
\end{align}
where 
\begin{subequations}
\begin{align}
m_{\hat{\pi}^0}^2 &= \frac{B}{2f_\pi} (m_u + m_d) , \\
m_{\pi_8}^2 &= \frac{B}{6 f_\pi} (m_u + m_d + 4 m_s), \\
m_{\pi_9}^2 &= \frac{B}{3 f_\pi} (m_u + m_d + m_s) + \frac{3}{2 C f_\pi^2}, \\
\bar{m}_{\hat{\phi}}^2 &= m_{\hat{\phi}}^2 + \frac{\tilde{\xi}_g^2}{C v^2}.
\end{align} \label{eq:meson-mass-parameters1}
\end{subequations} 
and
\begin{subequations}
\begin{align}
\delta &\equiv \frac{B}{2 f_\pi} (m_u - m_d), \\
\Delta &= \frac{B}{3 \sqrt{2} f_\pi} (m_u + m_d - 2 m_s), \\
\delta m_3^2 &= -\frac{B}{2 v} (\xi^p_u m_u - \xi^p_d m_d), \\
\delta m_8^2 &= -\frac{B}{2 \sqrt{3} v} (\xi^p_u m_u + \xi^p_d m_d - 2 \xi^p_s m_s), \\
\delta m_9^2 &= -\frac{B}{\sqrt{6} v} (\xi^p_u m_u + \xi^p_d m_d + \xi^p_s m_s) - \sqrt{\frac{3}{2}}\frac{\tilde{\xi}_g}{C f_\pi v}.
\end{align}\label{eq:meson-mass-parameters2}
\end{subequations} 
The matrix $O_\eta$ is the rotation matrix for the $\eta$--$\eta'$ mixing as
\begin{align}
O_\eta =
\begin{pmatrix}
1 & 0 & 0 & 0 \\
0 & \cos\theta_\eta & \sin\theta_\eta & 0 \\
0 & -\sin\theta_\eta & \cos\theta_\eta & 0 \\
0 & 0 & 0 & 1
\end{pmatrix}.
\end{align}
This mass term can be diagonalized by the orthogonal matrix $O$ as
\begin{align}
O^T M^2_{\rm mass} O = \mathrm{diag}(m_{\pi^0}^2, m_{\eta}^2, m_{\eta'}^2, m_{\phi}^2),    
\end{align}
where $\pi^0, \eta, \eta'$ and $\phi$ denote the mass eigenstates.

In our numerical analyses, we use the following values,
\begin{subequations}
\begin{align}
m_{\pi^\pm} &= 135~\mathrm{MeV},\quad  m_{\pi^0} = 135~\mathrm{MeV}, \\
m_{K^\pm}   &= 494~\mathrm{MeV},\quad m_{K^0} = 498~\mathrm{MeV}, \\
m_\eta      &= 548~\mathrm{MeV},\quad m_{\eta'} = 958~\mathrm{MeV},\quad \theta_\eta = -13.0^\circ.
\end{align}
\end{subequations}
Using these numbers, we obtain $\delta = - (63.0)^2$~MeV$^2$ which is one order of magnitude smaller than the above meson masses. When we ignore $\delta$, the diagonalization matrix $O$ can be obtained analytically \cite{Domingo:2016yih}. The leading-order terms of the mixing with $\hat{\phi}$ are given by
\begin{subequations}
\begin{align}
O_{\hat{\phi} \pi^0} &= \frac{\delta m_3^2}{m_{\pi}^2 - \bar{m}_{\phi}^2}, \\
O_{\hat{\phi} \eta} &= \frac{c_\eta \delta m_8^2 - s_\eta \delta m_9^2}{m_{\eta}^2 - m_{\phi}^2 }, \\
O_{\hat{\phi} \eta'} &= \frac{c_\eta \delta m_9^2 +  s_\eta \delta m_8^2}{m_{\eta'}^2 - m_{\phi}^2 }, \\
O_{\hat{\phi} \phi} &= 1,
\end{align}
\end{subequations}
where $\sin\theta_\eta$ and $\cos\theta_\eta$ are denoted by $s_\eta$ and $c_\eta$, respectively.
This expression is useful to understand the mixing behavior. 
Note that we numerically calculate $O$ in our analysis.

\section{Meson interactions} \label{apdx:meson-couplings}
The coupling constants for meson self-interactions as well as the dark scalar and meson interactions with the dark CP violation are written as 
\begin{align}
\mathcal L_{\mathrm{meson,int}} \equiv C_{ij \phi} \pi_i \pi_j \phi + C_{ijk\phi} \pi_i \pi_j \pi_k \phi + C_{ijkl} \pi_i \pi_j \pi_k \pi_l,
\end{align}
where $i, j, k, l$ run over the mesons in their mass eigenstates. In the results presented below, the quark masses and $\delta$ can be expressed in terms of the meson masses using the relations in Eqs.~\eqref{eq:meson-mass-parameters1} and \eqref{eq:meson-mass-parameters2}.

\subsection{$\phi$--two-meson couplings}
The $\phi$--two-meson couplings are given by
\begin{subequations}
\begin{align}
\pi^0\pi^0\phi:\quad C_{\pi^0\pi^0\phi} &= -\frac{B\big(m_d\xi_d^s+m_u\xi_u^s\big)}{4f_\pi v}, \\
\pi^+\pi^-\phi:\quad C_{\pi^+\pi^-\phi} &= -\frac{B\big(m_d\xi_d^s+m_u\xi_u^s\big)}{2f_\pi v}, \\
\pi^0\eta\phi:\quad C_{\pi^0\eta\phi} &= \frac{B(\sqrt2c_\eta-2s_\eta)\big(m_d\xi_d^s-m_u\xi_u^s\big)}{2\sqrt3f_\pi v}, \\
K^+K^-\phi:\quad C_{K^+K^-\phi} &= -\frac{B\big(m_s\xi_s^s+m_u\xi_u^s\big)}{2f_\pi v}, \\
K^0\bar K^0\phi:\quad C_{K^0\bar K^0\phi} &= -\frac{B\big(m_d\xi_d^s+m_s\xi_s^s\big)}{2f_\pi v}, \\
\pi^0\eta'\phi:\quad C_{\pi^0\eta'\phi} &= \frac{B(2c_\eta+\sqrt2s_\eta)\big(m_d\xi_d^s-m_u\xi_u^s\big)}{2\sqrt6f_\pi v}, \\
\eta\eta\phi:\quad C_{\eta\eta\phi} &= -\frac{B\left[(\sqrt2c_\eta-2s_\eta)^2\big(m_d\xi_d^s+m_u\xi_u^s\big) +4(\sqrt2c_\eta+s_\eta)^2m_s\xi_s^s\right]}{24f_\pi v}, \\
\eta\eta'\phi:\quad C_{\eta\eta'\phi} &= -\frac{B(2\sqrt2c_{2\eta}-s_{2\eta}) \big(m_d\xi_d^s-2m_s\xi_s^s+m_u\xi_u^s\big)}{12f_\pi v}, \\
\eta'\eta'\phi:\quad C_{\eta'\eta'\phi} &= -\frac{B\left[(2c_\eta+\sqrt2s_\eta)^2\big(m_d\xi_d^s+m_u\xi_u^s\big) +4(c_\eta-\sqrt2s_\eta)^2m_s\xi_s^s\right]}{24f_\pi v}.
\end{align}
\end{subequations}

\subsection{$\phi$--three-meson couplings}
The $\phi$--three-meson couplings are given by
\begin{align}
a_\eta &\equiv \frac{c_\eta-\sqrt2s_\eta}{\sqrt6}, ~~~ b_\eta \equiv -\frac{\sqrt2c_\eta+s_\eta}{\sqrt3}.
\end{align}
\begin{subequations}
\begin{align}
\pi^0\pi^0\pi^0\phi:\quad C_{\pi^0\pi^0\pi^0\phi} &= \frac{B\big(m_d\xi_d^p-m_u\xi_u^p\big)}{12f_\pi^2 v}, \\
\pi^+\pi^-\pi^0\phi:\quad C_{\pi^+\pi^-\pi^0\phi} &= \frac{B\big(m_d\xi_d^p-m_u\xi_u^p\big)}{6f_\pi^2 v}, \\
\pi^0\pi^0\eta\phi:\quad C_{\pi^0\pi^0\eta\phi} &= -\frac{B a_\eta\big(m_d\xi_d^p+m_u\xi_u^p\big)}{2\sqrt2f_\pi^2 v}, \\
\pi^+\pi^-\eta\phi:\quad C_{\pi^+\pi^-\eta\phi} &= -\frac{B a_\eta\big(m_d\xi_d^p+m_u\xi_u^p\big)}{\sqrt2f_\pi^2 v}, \\
\pi^0K^+K^-\phi:\quad C_{\pi^0K^+K^-\phi} &= -\frac{B\big(m_s\xi_s^p+2m_u\xi_u^p\big)}{6f_\pi^2 v}, \\
\pi^0K^0\bar K^0\phi:\quad C_{\pi^0K^0\bar K^0\phi} &= \frac{B\big(2m_d\xi_d^p+m_s\xi_s^p\big)}{6f_\pi^2 v}, \\
\pi^+K^-K^0\phi:\quad C_{\pi^+K^-K^0\phi} &= -\frac{B\big(m_d\xi_d^p+m_s\xi_s^p+m_u\xi_u^p\big)}{3\sqrt2f_\pi^2 v}, \\
\pi^-K^+\bar K^0\phi:\quad C_{\pi^-K^+\bar K^0\phi} &= -\frac{B\big(m_d\xi_d^p+m_s\xi_s^p+m_u\xi_u^p\big)}{3\sqrt2f_\pi^2 v}, \\
\pi^0\pi^0\eta'\phi:\quad C_{\pi^0\pi^0\eta'\phi} &= \frac{B b_\eta\big(m_d\xi_d^p+m_u\xi_u^p\big)}{4f_\pi^2 v}, \\
\pi^0\eta\eta\phi:\quad C_{\pi^0\eta\eta\phi} &= \frac{B a_\eta^2\big(m_d\xi_d^p-m_u\xi_u^p\big)}{2f_\pi^2 v}, \\
\pi^+\pi^-\eta'\phi:\quad C_{\pi^+\pi^-\eta'\phi} &= \frac{B b_\eta\big(m_d\xi_d^p+m_u\xi_u^p\big)}{2f_\pi^2 v}, \\
K^+K^-\eta\phi:\quad C_{K^+K^-\eta\phi} &= -\frac{B\left[2m_u(2a_\eta+b_\eta)\xi_u^p+3m_s b_\eta\xi_s^p\right]} {6\sqrt2f_\pi^2 v}, \\
K^0\bar K^0\eta\phi:\quad C_{K^0\bar K^0\eta\phi} &= -\frac{B\left[2m_d(2a_\eta+b_\eta)\xi_d^p+3m_s b_\eta\xi_s^p\right]} {6\sqrt2f_\pi^2 v}, \\
\pi^0\eta\eta'\phi:\quad C_{\pi^0\eta\eta'\phi} &= -\frac{B a_\eta b_\eta\big(m_d\xi_d^p-m_u\xi_u^p\big)}{\sqrt2f_\pi^2 v}, \\
\eta\eta\eta\phi:\quad C_{\eta\eta\eta\phi} &= -\frac{B\left[(m_u\xi_u^p+m_d\xi_d^p)a_\eta^3+m_s\xi_s^p b_\eta^3\right]} {3\sqrt2f_\pi^2 v}, \\
K^+K^-\eta'\phi:\quad C_{K^+K^-\eta'\phi} &= \frac{B\left[m_s(-4a_\eta+b_\eta)\xi_s^p-2\sqrt2m_u(a_\eta-b_\eta)\xi_u^p\right]} {6\sqrt2f_\pi^2 v}, \\
K^0\bar K^0\eta'\phi:\quad C_{K^0\bar K^0\eta'\phi} &= \frac{B\left[m_s(-4a_\eta+b_\eta)\xi_s^p-2\sqrt2m_d(a_\eta-b_\eta)\xi_d^p\right]} {6\sqrt2f_\pi^2 v}, \\
\pi^0\eta'\eta'\phi:\quad C_{\pi^0\eta'\eta'\phi} &= \frac{B b_\eta^2\big(m_d\xi_d^p-m_u\xi_u^p\big)}{4f_\pi^2 v}, \\
\eta\eta\eta'\phi:\quad C_{\eta\eta\eta'\phi} &= \frac{B\left[\big(m_d\xi_d^p+m_u\xi_u^p\big)a_\eta^2b_\eta -6m_s\xi_s^p a_\eta b_\eta^2\right]}{2f_\pi^2 v}, \\
\eta\eta'\eta'\phi:\quad C_{\eta\eta'\eta'\phi} &= -\frac{B\left[\big(m_d\xi_d^p+m_u\xi_u^p\big)a_\eta b_\eta^2 +4m_s\xi_s^p a_\eta^2b_\eta\right]}{2\sqrt2f_\pi^2 v}, \\
\eta'\eta'\eta'\phi:\quad C_{\eta'\eta'\eta'\phi} &= \frac{B\left[\big(m_d\xi_d^p+m_u\xi_u^p\big)b_\eta^3 -8m_s\xi_s^p a_\eta^3\right]}{12f_\pi^2 v}.
\end{align}
\end{subequations}

\subsection{Quadratic-meson couplings}
The quadratic-meson couplings are given by
\begin{subequations}
\begin{align}
a_\eta \equiv \frac{c_\eta-\sqrt2s_\eta}{\sqrt6}, ~~~ b_\eta \equiv -\frac{\sqrt2c_\eta+s_\eta}{\sqrt3}, ~~~
a_{\eta'} \equiv -\frac{b_\eta}{\sqrt2},~~~ b_{\eta'} \equiv \sqrt2a_\eta.
\end{align}
\end{subequations}

\begin{subequations}
\begin{align}
\pi^0\pi^0\pi^0\pi^0:\quad C_{\pi^0\pi^0\pi^0\pi^0} &= \frac{B(m_u+m_d)}{48f_\pi^3}, \\
\pi^+\pi^-\pi^0\pi^0:\quad C_{\pi^+\pi^-\pi^0\pi^0} &= \frac{B(m_u+m_d)}{12f_\pi^3}, \\
\pi^0\pi^0\pi^0\eta:\quad C_{\pi^0\pi^0\pi^0\eta} &= \frac{B\sqrt2a_\eta(m_u-m_d)}{12f_\pi^3}, \\
\pi^+\pi^-\pi^0\eta:\quad C_{\pi^+\pi^-\pi^0\eta} &= \frac{B\sqrt2a_\eta(m_u-m_d)}{6f_\pi^3}, \\
\pi^0\pi^0K^+K^-:\quad C_{\pi^0\pi^0K^+K^-} &= \frac{B(3m_u+m_s)}{24f_\pi^3}, \\
\pi^0\pi^0K^0\bar K^0:\quad C_{\pi^0\pi^0K^0\bar K^0} &= \frac{B(3m_d+m_s)}{24f_\pi^3}, \\
\pi^+\pi^0K^-K^0:\quad C_{\pi^+\pi^0K^-K^0} &= \frac{B\sqrt2(m_u-m_d)}{24f_\pi^3}, \\
\pi^-\pi^0K^+\bar K^0:\quad C_{\pi^-\pi^0K^+\bar K^0} &= \frac{B\sqrt2(m_u-m_d)}{24f_\pi^3}, \\
\pi^+\pi^-K^0\bar K^0:\quad C_{\pi^+\pi^-K^0\bar K^0} &= \frac{B(m_u+2m_d+m_s)}{12f_\pi^3}, \\
\pi^0\pi^0\pi^0\eta':\quad C_{\pi^0\pi^0\pi^0\eta'} &= \frac{B\sqrt2a_{\eta'}(m_u-m_d)}{12f_\pi^3}, \\
\pi^0\pi^0\eta\eta:\quad C_{\pi^0\pi^0\eta\eta} &= \frac{Ba_\eta^2(m_u+m_d)}{4f_\pi^3}, \\
\pi^+\pi^-\pi^0\eta':\quad C_{\pi^+\pi^-\pi^0\eta'} &= \frac{B\sqrt2a_{\eta'}(m_u-m_d)}{6f_\pi^3}, \\
\pi^+\pi^-\eta\eta:\quad C_{\pi^+\pi^-\eta\eta} &= \frac{Ba_\eta^2(m_u+m_d)}{2f_\pi^3}, \\
\pi^0K^+K^-\eta:\quad C_{\pi^0K^+K^-\eta} &= \frac{B\sqrt2\left[(3a_\eta+b_\eta)m_u+(a_\eta+b_\eta)m_s\right]}{12f_\pi^3}, \\
\pi^0K^0\bar K^0\eta:\quad C_{\pi^0K^0\bar K^0\eta} &= -\frac{B\sqrt2\left[(3a_\eta+b_\eta)m_d+(a_\eta+b_\eta)m_s\right]}{12f_\pi^3}, \\
\pi^+K^-K^0\eta:\quad C_{\pi^+K^-K^0\eta} &= \frac{B\left[(3a_\eta+b_\eta)(m_u+m_d) +2(a_\eta+b_\eta)m_s\right]}{12f_\pi^3}, \\
\pi^-K^+\bar K^0\eta:\quad C_{\pi^-K^+\bar K^0\eta} &= \frac{B\left[(3a_\eta+b_\eta)(m_u+m_d) +2(a_\eta+b_\eta)m_s\right]}{12f_\pi^3}, \\
\pi^0\pi^0\eta\eta':\quad C_{\pi^0\pi^0\eta\eta'} &= \frac{Ba_\eta a_{\eta'}(m_u+m_d)}{2f_\pi^3}, \\
\pi^0\eta\eta\eta:\quad C_{\pi^0\eta\eta\eta} &= \frac{B\sqrt2a_\eta^3(m_u-m_d)}{6f_\pi^3}, \\
\pi^+\pi^-\eta\eta':\quad C_{\pi^+\pi^-\eta\eta'} &= \frac{Ba_\eta a_{\eta'}(m_u+m_d)}{f_\pi^3}.
\end{align}
\end{subequations}

\begin{subequations}
\begin{align}
\pi^0K^+K^-\eta':\quad C_{\pi^0K^+K^-\eta'} &= \frac{B\sqrt2\left[(3a_{\eta'}+b_{\eta'})m_u+(a_{\eta'}+b_{\eta'})m_s\right]}{12f_\pi^3}, \\
\pi^0K^0\bar K^0\eta':\quad C_{\pi^0K^0\bar K^0\eta'} &= -\frac{B\sqrt2\left[(3a_{\eta'}+b_{\eta'})m_d+(a_{\eta'}+b_{\eta'})m_s\right]}{12f_\pi^3}, \\
\pi^+K^-K^0\eta':\quad C_{\pi^+K^-K^0\eta'} &= \frac{B\left[(3a_{\eta'}+b_{\eta'})(m_u+m_d) +2(a_{\eta'}+b_{\eta'})m_s\right]}{12f_\pi^3}, \\
\pi^-K^+\bar K^0\eta':\quad C_{\pi^-K^+\bar K^0\eta'} &= \frac{B\left[(3a_{\eta'}+b_{\eta'})(m_u+m_d) +2(a_{\eta'}+b_{\eta'})m_s\right]}{12f_\pi^3}, \\
\pi^0\pi^0\eta'\eta':\quad C_{\pi^0\pi^0\eta'\eta'} &= \frac{Ba_{\eta'}^2(m_u+m_d)}{4f_\pi^3}, \\
\pi^0\eta\eta\eta':\quad C_{\pi^0\eta\eta\eta'} &= \frac{B\sqrt2a_\eta^2a_{\eta'}(m_u-m_d)}{2f_\pi^3}, \\
\pi^0\eta\eta'\eta':\quad C_{\pi^0\eta\eta'\eta'} &= \frac{B\sqrt2a_\eta a_{\eta'}^2(m_u-m_d)}{2f_\pi^3}.
\end{align}
\end{subequations}

\vspace*{-8mm}
\bibliography{biblio}

\end{document}